\documentclass[a4,12pt]{article}
\usepackage{graphicx}
\usepackage{amsmath,amssymb}
\usepackage{ulem}

\usepackage{color}
\usepackage{here}

\begin{document}
\setlength{\baselineskip}{20pt}

\makeatletter
 \renewcommand{\theequation}{%
  \thesection.\arabic{equation}}
 \@addtoreset{equation}{section}
\makeatother

\newtheorem{remark}{Remark}[section]
\newtheorem{theo}{Theorem}[section]
\newtheorem{lemma}{Lemma}[section]
\newtheorem{prop}{Proposition}[section]
\newtheorem{assumption}{Assumption}[section]
\newtheorem{cor}{Corollary}[section]
\newtheorem{example}{Example}[section]

\def\ep{\varepsilon}
\def\ov{\overline}
\def\del{\partial}
\def\norm{\parallel}
\def\no{\noindent}
\def\lam{\lambda}
\def\dis{\displaystyle}
\def\bhat{\widehat}
\def\lap{\bigtriangleup}
\def\R{\mbox{\bf R}}
\def\lg{\langle}
\def\rg{\rangle}

\newcommand{\Qed}{ $\square$ }

\title{
Oscillations and Bifurcation Structure of Reaction-Diffusion Model for Cell Polarity Formation
}

\author{
Masataka Kuwamura\thanks{
Graduate School of Human Development and Environment,
Kobe University, Kobe 
657-8501, Japan,
email: kuwamura@main.h.kobe-u.ac.jp }\;
,
Hirofumi Izuhara\thanks{
Faculty of Engineering, 
University of Miyazaki, 
1-1 Gakuen Kibanadai-nishi, Miyazaki, 889-2192, Japan, 
email: izuhara@cc.miyazaki-u.ac.jp}\,
and 
Shin-ichiro Ei\thanks{
Department of Mathematics,
Hokkaido University, Sapporo 
060-0810, Japan,
email: Eichiro@math.sci.hokudai.ac.jp}
}

\date{}


\maketitle
\begin{abstract}
We investigate the oscillatory dynamics and bifurcation structure
of a reaction-diffusion system with bistable nonlinearity and mass conservation,
which was proposed by [Otsuji et al,  PLoS Comp. Biol. 3 (2007), e108].
The system is a useful model for understanding cell polarity formation.
We show that this model exhibits four different spatiotemporal patterns
including two types of oscillatory patterns, which can be regarded as 
cell polarity oscillations with the reversal and non-reversal of polarity, respectively.
The trigger causing these patterns is a diffusion-driven (Turing-like) instability. 
Moreover, we investigate the effects of extracellular signals on the
cell polarity oscillations. 
\\

\no
\\


\no


\end{abstract}


\section{Introduction }

Reaction-diffusion systems with bistable nonlinearity 
are useful for understanding
basic biological phenomena such as
the population dynamics of competing species and propagating pulses on the nerve axon
\cite{Mu, O}.
As for cell polarization, which is a key process in cell division and differentiation, 
\cite{MJE1} proposed a reaction-diffusion model with bistable nonlinearity
for understanding the wave-pinning phenomena: A transient and localized stimulus to 
an unpolarized cell is spatially amplified to result in a robust subdivision of the cell into
two clearly defined regions, front and back.  
Although the bistability in their model, which is represented by a Hill function, 
may not explicitly correspond to concrete molecular networks, it 
is a simple (phenomenological) model and 
hence allows us to explain a mechanism refereed to as the wave-pinning 
by a mathematical analysis.
Moreover, \cite{MJE} investigated the bifurcation structure of the model in \cite{MJE1}.

The maintenance of cell polarity can be considered as a dynamic process.
In fact, cells in a developing epithelial tissue alternately repeat
the establishment and loss of polarity in a cell cycle-dependent manner.
This repetition is known as a cell polarity oscillation \cite{DR},
where the reversal of polarity does not occur in such a way that
the cells in an epithelial tissue are organized into layers.
Moreover, cell polarity oscillations are also observed in cell migration, which 
is the directed movement of a single cell or a group of cells in response to chemical 
and/or mechanical signals. 
For example, bacteria cells during rippling (accordion-like movements) \cite{GM, SN} and
migrating melanoma cells derived from tumors \cite{PH} can reverse their polarity in an oscillatory fashion.

For understanding cell polarity oscillations, some
mathematical models have been proposed.
For example, \cite{PH} proposed an ODE system based on biomolecular dynamics at the
front and back lamellipodia in a migration cell to better understand their experimental results 
concerning the change of cell polarity in melanoma cells derived from tumors.
\cite{GM} proposed an ODE system based on three proteins to understand
cell polarity oscillations with the polarity reversal of
bacteria cells during rippling (accordion-like movements) \cite{SN}.
\cite{LL} proposed a delay reaction-diffusion equation with a nonlocal term for
studying the effects of a delayed negative feedback on cell polarity oscillations with the aid of numerical simulations. 
Recently, \cite{TW} generalized the ODE system in \cite{GM}, and proposed an ODE system   
that generates four different mechanisms for switching cell polarity,
including a mechanism to generate the periodic reversal of polarity. 

In this paper, we consider a reaction-diffusion model with bistable nonlinearity 
proposed by \cite{Ot},
which exhibits dynamics similar to that of considerably realistic reaction-diffusion models 
based on concrete molecular species and networks 
concerning cell polarity formation \cite{NSL, Ot, SN2}:

\begin{equation}\label{a1}
\left \{
\begin{array}{rcl}
\dot{u} & = & d_1 u_{xx} +  f(u,v)  \\[1ex]
\dot{v} & = & d_2 v_{xx}  -  f(u,v)
\end{array} 
\right.
\end{equation}
with 
\begin{equation}\label{a2}
f(u, v) = a_1 ( u + v) \{ (\alpha u + v) (u + v) - a_2 \}
\end{equation}
and
\begin{equation}\label{a2x}
\alpha = \frac{ d_1}{d_2},
\end{equation}
where $d_1, d_2, a_1$, and $a_2$ are positive constants.
We consider that 
$u$ and $v$ correspond to chemicals in the membrane and cytosol of a cell, 
respectively. Since
the diffusion in the cytosol is faster than that in the membrane,
the diffusion coefficients $d_1$ and $d_2$ satisfy the condition
\[
d_1 < d_2,
\]
which implies $0 < \alpha < 1$.
We note that $u$ and $v$ should be considered as deviations from the reference values
of the concentrations of the chemicals because either $u$ or $v$ can be negative, even if
the initial values of $u$ and $v$ are positive as seen in the next section.
Moreover, by setting $a_2 \to a_2 - \ep a_E(x)$ in \eqref{a2},
we obtain the perturbed system of \eqref{a1}, i.e.,
\begin{equation}\label{a4}
\left \{
\begin{array}{rcl}
\dot{u} & = & d_1 u_{xx} + \{  f(u,v) + \ep g(x, u, v) \} \\[1ex]
\dot{v} & = & d_2 v_{xx}  - \{ f(u,v) + \ep g(x, u, v)  \}
\end{array} 
\right.
\end{equation}
with
\begin{equation}\label{a5}
g(x, u, v) =   a_1 ( u + v) a_E(x),
\end{equation}
where $a_E(x)$ is a smooth function that
represents the waveform of an extracellular signal, and  $\ep$ is a small positive parameter that
controls the amplitude of the extracellular signal.
We consider \eqref{a1} and \eqref{a4}
on an interval $I = ( -K/2, K/2)$ under the periodic boundary condition
for sufficiently large $K$.

In general, a reaction-diffusion system \eqref{a1} with a smooth function $f$ is called a 
reaction-diffusion system with 
mass conservation because 
\begin{equation}\label{ax2}
\xi := \frac{1}{K} \int_0^K \left( u(x, 0) + v(x, 0) \right) dx \equiv 
\frac{1}{K} \int_0^K \left( u(x, t) + v(x, t) \right) dx 
\end{equation}
holds for any (smooth) solutions under the periodic or Neumann boundary conditions.
For certain functions $f$ and boundary conditions,
some mathematical aspects of \eqref{a1}
have been studied elsewhere \cite{CMS, Is, JM, MJE, MO, Ok}, and one of the typical dynamics of \eqref{a1}
is summarized as follows:
Solutions with initial values near a homogeneous equilibrium destabilized 
through the same mechanism as the Turing instability
eventually approach a localized unimodal stationary pattern (see, Figure~\ref{zu1}).
This dynamics is interpreted as cell polarization, and the position of the peak of 
the localized unimodal pattern determines the direction of the polarity.
As for the function $f$ given by \eqref{a2}, it was numerically shown in \cite{Ot} that 
\eqref{a1} has such dynamics under certain conditions. 
In addition, \cite{Ot} performed a formal stability analysis for the final localized unimodal stationary pattern of \eqref{a1}, and
numerically showed that a solution of \eqref{a4} starting from  
the final pattern of \eqref{a1} can translationally move to the maximum point of $a_E(x)$
when $a_E(x)$ is given by $a_E(x) = A \cos ( 2\pi x /K)$.
This implies that cell polarity can be controlled by extracellular signals.

In contrast, using the normal form theory, it was shown in \cite{Ok} that 
\eqref{a1} can have a unimodal stationary pattern, which changes its stability due to a Hopf bifurcation 
when we vary $d_1$ as a bifurcation parameter. Consequently, 
\eqref{a1} with \eqref{a2} has a spatially nonhomogeneous limit cycle with small amplitude under certain conditions.

The purpose of this paper is to show that
\eqref{a1} with \eqref{a2} exhibits four different spatiotemporal patterns
including two types of oscillatory patterns, which can be regarded as 
cell polarity oscillations with the reversal and non-reversal of polarity, respectively.
Our numerical simulations show that \eqref{a1} with \eqref{a2}
has large amplitude limit cycles (see, Figure~\ref{zu3}).
These limit cycles exhibit an alternating repetition of unimodal and spatially homogeneous patterns, 
and hence can be interpreted as cell polarity oscillations. Moreover, 
using a numerical bifurcation analysis, 
we investigate the bifurcation structure of \eqref{a1} with \eqref{a2}, which generates 
the limit cycles when we vary $d_1$ as a bifurcation parameter. 
The result suggests that $d_1$ and $\xi$ defined by \eqref{ax2} play a 
key role in the switching cell polarity.
Furthermore, when $a_E(x)$ is given by $a_E(x) = A \cos ( 2\pi x /K)$,
we investigate the dynamics of  \eqref{a4} with \eqref{a2} and \eqref{a5}.
In particular, we numerically show that \eqref{a4} with \eqref{a2} and \eqref{a5}
has limit cycles exhibiting such an alternating repetition. 
To understand the controllability of cell polarity oscillations by extracellular signals,
we examine the position of the peak of the periodically appearing unimodal patterns.

The remainder of this paper is organized as follows. In Section 2, 
we describe the bistability of an ODE system obtained from \eqref{a1} by dropping 
the diffusion terms. Our characterization of the bistability is different from that of
\cite{MJE1, MJE}. In Section 3,
we investigate spatiotemporal patterns of the unperturbed system \eqref{a1} with \eqref{a2}.
We give a concrete expression of a
localized unimodal stationary pattern of \eqref{a1} with \eqref{a2} for sufficiently large $K$.
As a prominent feature, this model enables us to perform
concrete calculations for a mathematical analysis. 
Moreover, we numerically show the formation process of the localized unimodal stationary pattern
and the existence of limit cycles, which exhibit
an alternating repetition of unimodal and spatially homogeneous patterns.
Consequently, we see that
\eqref{a1} with \eqref{a2} exhibits four different spatiotemporal patterns
including two types of oscillatory patterns, which can be regarded as 
cell polarity oscillations with the reversal and non-reversal of polarity, respectively.
In Section 4, using a numerical bifurcation analysis, we investigate the bifurcation structure of \eqref{a1} with \eqref{a2}
when we vary $d_1$ as a bifurcation parameter.
Our numerical bifurcation analysis shows that 
the limit cycles exhibiting the alternating repetition bifurcate from the localized unimodal
stationary pattern through a Hopf bifurcation, and that the limit cycles become unstable 
through a period doubling bifurcation. 
These results support the results obtained by numerical simulations in Section 2, and show that
the trigger causing the spatiotemporal patterns is a diffusion-driven (Turing-like) instability. 
In Section 5, we investigate the dynamics of spatiotemporal patterns 
in the perturbed system \eqref{a4} with \eqref{a2} and \eqref{a5} 
when $a_E(x)$ is given by $a_E(x) = A \cos ( 2\pi x /K)$.
We analytically derive the equation of motion for the translational movement of the localized unimodal pattern, which gives a
mathematical justification of the results in \cite{Ot}. 
Moreover, we show that \eqref{a4} with \eqref{a2} and \eqref{a5} has
limit cycles exhibiting such an alternating repetition, and 
examine whether
the position of the peak of the periodically appearing unimodal pattern
can be determined by the maximum point of $a_E(x)$.
These results are useful for understanding the effects of extracellular signals on
cell polarity oscillations.
Section 6 discusses the results of this study.

\section{ The dynamics of ODE system }

In this section, 
we investigate the dynamics of an ODE system obtained from \eqref{a1} by dropping 
the diffusion terms. 
Our characterization of the bistability of the ODE system is different from that of
\cite{MJE1, MJE}.

\begin{figure}
\begin{center}
\resizebox{8cm}{8cm}{\includegraphics{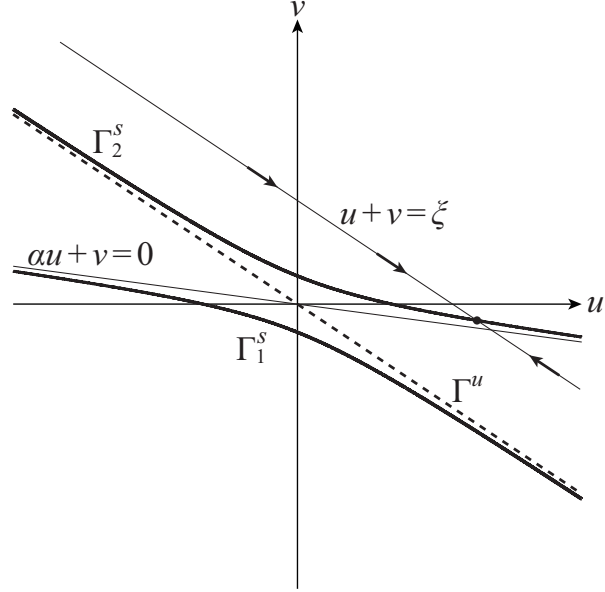}} 
\caption[]{ \label{zu1xy}
The dynamics of the ODE system \eqref{aaa1} with \eqref{aaa2}.
}
\end{center}
\end{figure}

We consider the dynamics of 
\begin{equation}\label{aaa1}
\left \{
\begin{array}{rcl}
\dot{u} & = &   f(u,v)  \\[1ex]
\dot{v} & = & - f(u,v)
\end{array} 
\right.
\end{equation}
with
\begin{equation}\label{aaa2}
f(u, v) = a_1 ( u + v) \{ (\alpha u + v) (u + v) - a_2 \},
\end{equation}
where $a_1 > 0 $, $a_2 > 0$ and $0 < \alpha < 1$.
It is easy to see that $f(u , v )=0$, an equation for $u$ with arbitrarily fixed $v$,  
has three roots $h_1(v) < -v < h_2(v)$, where
\[
\dis\lim_{v \to \infty} \big{|}   h_1(v) + \frac{v}{\alpha} \, \big{|} = 0 \ \ \text{and} \ \ 
\lim_{v \to -\infty} | h_1(v) + v \, | = 0,
\]
and
\[
\dis\lim_{v \to \infty} | h_2(v) + v \, | = 0  \ \ \text{and} \ \ 
\lim_{v \to -\infty} \big{|} h_2(v) + \frac{v}{\alpha} \, \big{|} = 0.
\]
We note that any solution of \eqref{aaa1} satisfies $u + v = \xi $ for some $\xi$, and that
\begin{equation}\label{aaa3}
\dot{u} = f(u, \xi - u) = - a_1 \xi^2 ( 1 - \alpha) u + a_1 \xi^3 - a_1 a_2 \xi
\end{equation}
has a unique stable equilibrium for $\xi \neq 0$ by $0 < \alpha < 1$. Hence, we see that 
\eqref{aaa1} has a stable equilibrium 
\begin{equation}\label{aaaa4}
\left( \frac{\xi^2 - a_2 }{(1- \alpha)\xi}, \frac{a_2 - \alpha\xi^2}{(1-\alpha)\xi} \right) 
\end{equation}
with $\xi \neq 0$, which corresponds to the stable equilibrium of \eqref{aaa3} for
$\xi \neq 0$. Therefore, we see that
\eqref{aaa1} has two families of stable equilibria 
$\Gamma_1^s = \{ \, ( h_1(v), v) \, | \, v \in \text{\bf R} \, \}$ and 
$\Gamma_2^s = \{ \, ( h_2(v), v) \, | \, v \in \text{\bf R} \, \}$, and 
a family of unstable equilibria $\Gamma^u = \{ \, ( -v, v) \, | \, v \in \text{\bf R} \, \}$.
Thus, we can consider that \eqref{aaa1} is bistable in the usual sense because 
it has two equivalent stable states $\Gamma_1^s$ and $\Gamma_2^s$, 
and one unstable state $\Gamma^u$,
which is a separatrix as seen in Figure~\ref{zu1xy}.
Moreover, we note that the dynamics of \eqref{aaa1} does not qualitatively change when we vary 
$\alpha \in (0, 1)$.

\begin{remark}\label{rem11} \rm
Although we cannot apply a standard argument concerning 
the invariant region of reaction-diffusion systems \cite{Smo}, 
the phase plane in Figure~\ref{zu1xy} and numerical simulations suggest
 that $\{ \, (u, v)  \, | \, u + v > 0 \, \}$ is an invariant region of the PDE
\eqref{a1} with \eqref{a2}
as well as the case of the ODE \eqref{aaa1} with \eqref{aaa2}.
\end{remark}

We note that for each fixed $v$, the ODE $\dot{u} = f( u, v)$ is monostable because
it has two unstable equilibria $h_1(v)$ and $h_2(v)$, and 
one stable equilibrium $-v$. This is equivalent to a condition in which
for each fixed $u$, the ODE $\dot{v} = - f( u, v)$ is bistable because
it has two stable equilibria $h_1^{-1}(u)$ and $h_2^{-1}(u)$, and 
one unstable equilibrium $-u$. 
The condition in which $\dot{u} = f( u, v)$ is monostable for each fixed $v$ is 
opposite to a condition in \cite{MJE1, MJE} in which
$\dot{u} = f( u, v)$ is bistable for each fixed $v$, 
which is used for a standard argument concerning the traveling waves and 
transition layers of reaction-diffusion systems (cf. \cite{NI} and the references therein). 
In contrast, we are concerned with unimodal patterns of 
\eqref{a1} with \eqref{a2}, which do not need such an argument.

\section{ Spatiotemporal patterns of unperturbed system}

In this section, 
we investigate spatiotemporal patterns of the unperturbed system \eqref{a1}
with \eqref{a2},
which is useful for understanding spontaneous cell polarization without extracellular signals. 
In particular, we numerically show 
limit cycles that exhibit
an alternating repetition of unimodal and spatially homogeneous patterns.
These limit cycles can be interpreted as cell polarity oscillations. 

\subsection{ Concrete expression of localized unimodal stationary pattern }

We give a concrete expression of a
localized unimodal stationary pattern of \eqref{a1} with \eqref{a2} for sufficiently large $K$.

We consider stationary solutions of \eqref{a1}, i.e.,
\begin{equation}\label{b1}
\left \{
\begin{array}{l}
d_1 u_{xx} +  f(u,v) = 0   \\[1ex]
d_2 v_{xx}  -  f(u,v) = 0,
\end{array} 
\right.
\end{equation}
where $f(u, v)$ is given by \eqref{a2}. Applying a transformation of variables
\begin{equation}\label{b2}
p:= u+v \ \ \ \ \text{and}  \ \ \ \ 
q:= \alpha u + v
\end{equation}
to \eqref{b1}, we have 
\begin{equation}\label{b3}
\left \{
\begin{array}{l}
q_{xx} = 0   \\[1ex]
d_1 d_2 p_{xx} + (d_2 - d_1) a_1 p(pq- a_2)=0.
\end{array} 
\right.
\end{equation}
Since $q$ satisfies the periodic boundary condition, it follows from 
the first equation of \eqref{b3} that $q(x) \equiv q_0$. 
Therefore, by the second equation of \eqref{b3}, we have
\begin{equation}\label{b4}
p_{xx} + c_1 p ( p - c_2 ) =0,
\end{equation}
where 
\[
c_1 =  \dis\frac{(d_2 - d_1)a_1 q_0}{d_1 d_2}  \ \ \ \ \text{and} \ \ \ \  
c_2 = \dis\frac{a_2}{q_0}.
\]
On the other hand, a direct calculation shows that
\[
p(x) = p_0 \text{sech}^2(bx)
\]
satisfies \eqref{b4} for
\begin{equation}\label{b7}
p_0 = \dis\frac{3a_2}{2q_0} 
\ \ \ \ \text{and} \ \ \ \  
b = \dis\frac{1}{2} \sqrt{ \dis\frac{(d_2 - d_1)a_1 a_2}{d_1 d_2}   }.
\end{equation}
Therefore, noting \eqref{b2}, we see that 
\begin{equation}\label{b8}
u = \varphi_1(x)
:= \dis\frac{1}{d_2 - d_1} 
\left( d_2 p_0 \text{sech}^2(bx) - \dis\frac{3a_2 d_2}{2 p_0}  \right)
\end{equation}
and
\begin{equation}\label{b8x}
v =
\varphi_2(x)
:=
\dis\frac{1}{d_2 - d_1} 
\left(
-d_1 p_0 \text{sech}^2(bx) + \dis\frac{3a_2 d_2}{2 p_0}  
\right)
\end{equation}
satisfy \eqref{b1}. 
Since $\text{sech}(bx) = O(e^{-b|x|})$ as $|x| \to \infty$,
it is easy to see that \eqref{a1} has a stationary solution 
$(\bar{u}(x), \bar{v}(x))$ satisfying the periodic boundary condition at $x = \pm K/2$, where 
\begin{equation}\label{b9}
\bar{u}(x) = \varphi_1(x) +  O(e^{-bK})
\end{equation}
and 
\begin{equation}\label{b10}
\bar{v}(x) = \varphi_2(x) +  O(e^{-bK})
\end{equation}
as $K \to \infty$. 
We consider that $(\bar{u}(x), \bar{v}(x))$ is a 
localized unimodal stationary pattern as shown in Figure~\ref{zu1}(b)
when $p_0 > 0$ and $K$ is sufficiently large. 
Moreover, we note that  
$(\bar{u}(x-c), \bar{v}(x-c))$ for some $c \in [-K/2, K/2)$ is 
also a stationary solution of \eqref{a1} because of the periodic boundary condition. 

It follows from \eqref{b9} and \eqref{b10} that
\begin{equation}\label{b11}
\bar{u}(x) + \bar{v}(x) = p_0 \text{sech}^2(bx) + O(e^{-bK})
\end{equation}
and
\begin{equation}\label{b11xx}
\alpha \bar{u}(x) + \bar{v}(x) = q_0  + O(e^{-bK}).
\end{equation}
Since
$$
\int_I \text{sech}^2(bx) dx = \int_{-\infty}^{\infty} \text{sech}^2(bx) dx + O(e^{-bK})
= \dis\frac{2}{b} +  O(e^{-bK}),
$$
integrating \eqref{b11} over $I=(-K/2, K/2)$, we have
\begin{equation}\label{b12}
p_0 = \dis\frac{b}{2} \int_I ( \bar{u}(x) + \bar{v}(x) ) dx + O(Ke^{-bK})
\end{equation}
as $K \to \infty$. This implies that the amplitude of the localized stationary pattern
given by \eqref{b9} and \eqref{b10}
can be explicitly determined by the total mass of components.

\subsection{Formation of localized unimodal pattern}

We numerically investigate 
the formation process of a localized unimodal pattern of \eqref{a1}
with \eqref{a2}.
Our numerical simulations show that 
a solution starting from 
a spatially homogeneous initial state with a small disturbance converges to 
$(\bar{u}(x-c), \bar{v}(x-c))$ for some $c \in [-K/2, K/2)$
under certain conditions, where $\bar{u}$ and $\bar{v}$ are given by \eqref{b9}
and \eqref{b10}, respectively.

Noting the mass conservation property \eqref{ax2} and \eqref{b12}, we see that
the amplitude of the localized stationary pattern given by \eqref{b9}
and \eqref{b10} is determined by 
\begin{equation}\label{b12xx}
p_0 = \dis\frac{bK}{2} \xi + O(Ke^{-bK})
\end{equation}
for sufficiently large $K$, where 
\begin{equation}\label{b12yy}
\xi = \dis\frac{1}{K} \int_I ( u(x, 0) + v(x, 0) ) dx
\end{equation}
can be considered as a parameter.
For each fixed $\xi$, 
we numerically solve \eqref{a1} for various values of $d_1$  
under the parameter values 
\begin{equation}\label{b13}
a_1 = 0.5, \ \ a_2 = 2.2, \ \ d_2 = 1.0.
\end{equation}
We choose $K = 10.0$ in such a way that $(\varphi_1(x), \varphi_2(x))$ can be numerically regarded as a
stationary solution satisfying the periodic boundary condition.
The initial value $(u(x, 0), v(x, 0))$ is given by 
$(\xi/2, \xi/2)$ with a small random perturbation,
which means that the initial states of $u$ and $v$ are almost the same 
and spatially homogeneous.  
Notice that these parameter values are the same as those in \cite{Ot}.

\begin{figure}
\begin{center}
\resizebox{13cm}{5cm}{\includegraphics{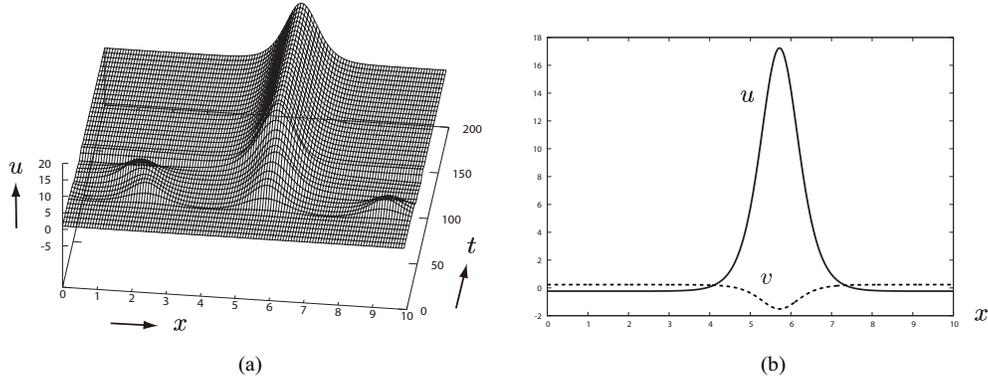}} 
\caption[]{ \label{zu1}
Formation of a localized unimodal stationary pattern for $d_1 = 0.1$ when $\xi = 2.0$.
The initial value $(u(x, 0), v(x, 0))$ is given by 
$(\xi/2, \xi/2)$ with a small random perturbation. 
(a) The dynamics of $u$-component.
The values of $u(x,t)$ on $0 \leq x \leq 10$ and $0 \leq t \leq 200$ are 
represented by a 3D graph. 
The profile of $v(x, t)$ is omitted here because the amplitude and spatial variation of $v(x, t)$ for each $t$
are relatively small as compared to those of $u(x, t)$. 
(b) The profiles of $u(x,t)$ and $v(x,t)$ for $t = 200$, 
which can be regarded as the localized unimodal stationary pattern given by
$(\bar{u}(x-c), \bar{v}(x-c))$ for some  $c \in [-K/2, K/2)$.
The rigid and dashed lines represent $u$- and $v$-components, respectively.
The numerical solutions on $0 \leq x \leq 10$ presented in this figure can be regarded as
those on $-5 \leq x \leq 5$ under the periodic boundary condition. 
It should be noted that we cannot predict the value of $c$,
the position of the peak of the localized unimodal pattern
in $u$-component, because of a small random perturbation to the initial value of a solution.
}
\end{center}
\end{figure}

We use a numerical scheme in \cite{KM} based on the pseudospectral method;
we obtain numerical solutions of \eqref{a1} on $0 \leq x \leq K$, 
which can be regarded as those on $-K/2 \leq x \leq K/2$ under the periodic boundary condition. 
The dynamics of numerical solutions consists of two stages. 
In the first stage, it is governed by the ODE \eqref{aaa3}, and 
the solutions approach a stable equilibrium of the ODE \eqref{aaa1}, which is given by
\eqref{aaaa4}.
Figure~\ref{zu1} shows the second stage where the solutions converge to the localized unimodal stationary pattern
given by $(\bar{u}(x-c), \bar{v}(x-c))$ for some  $c \in [-K/2, K/2)$
under the parameter values \eqref{b13} for $d_1 = 0.1$ when $\xi = 2.0$.
Notice that we cannot predict the value of $c$, 
the position of the peak of the localized unimodal stationary pattern, 
because of a small random perturbation to the initial value of a solution.
The formation process of a localized unimodal stationary pattern as presented in Figure~\ref{zu1}
can be interpreted as cell polarization \cite{Ot}.

Our simulations show that when $\xi = 2.0$, numerical solutions converge to a (localized) unimodal stationary pattern for 
$d_1 \lesssim 0.73$, where the accuracy of approximation for the (localized) unimodal pattern 
by $(\bar{u}(x-c), \bar{v}(x-c))$ becomes lower as $d_1$ increases. 
We can obtain similar results when $\xi \gtrsim 0.68$.

\begin{remark}\label{rem2} \rm
Numerical simulations for the large values of $\xi$ are delicate task.
In fact, when $\xi = 2.0$, 
the numerical result shown in Figure~\ref{zu1} can be obtained if we adopt the time step as $\Delta t = 0.02$. 
In contrast, when $\xi = 6.0$, 
we can obtain a correct numerical result as shown in Figure~\ref{zu1} if $\Delta t = 0.01$.
\end{remark}

\begin{remark}\label{rem2xyz} \rm
We perform numerical simulations for $\xi > 0$.
When $\xi < 0$, noting $p_0 < 0$ by \eqref{b12xx}, it follows from
\eqref{b8} and \eqref{b8x} that $(\bar{u}(x-c), \bar{v}(x-c))$
represents a localized anti-unimodal stationary pattern where
the $u$-component has a single hole. We can observe 
the dynamics of anti-unimodal patterns, which is similar to that of unimodal patterns
as shown in Figures~\ref{zu1} and \ref{zu3}. We do not consider the case $\xi = 0$ that 
numerical solutions do not move.
\end{remark}

\subsection{Spatiotemporal oscillatory patterns}

We cannot observe that
numerical solutions converge to the localized unimodal stationary pattern given by
$(\bar{u}(x-c), \bar{v}(x-c))$ under certain conditions.
In fact, when $\xi \lesssim 0.66$, our simulations show that 
the localized unimodal stationary pattern becomes unstable and 
spatiotemporal oscillatory patterns appear for some values of $d_1$ under the parameter values \eqref{b13}.

\begin{figure}
\begin{center}
\resizebox{125mm}{90mm}{\includegraphics{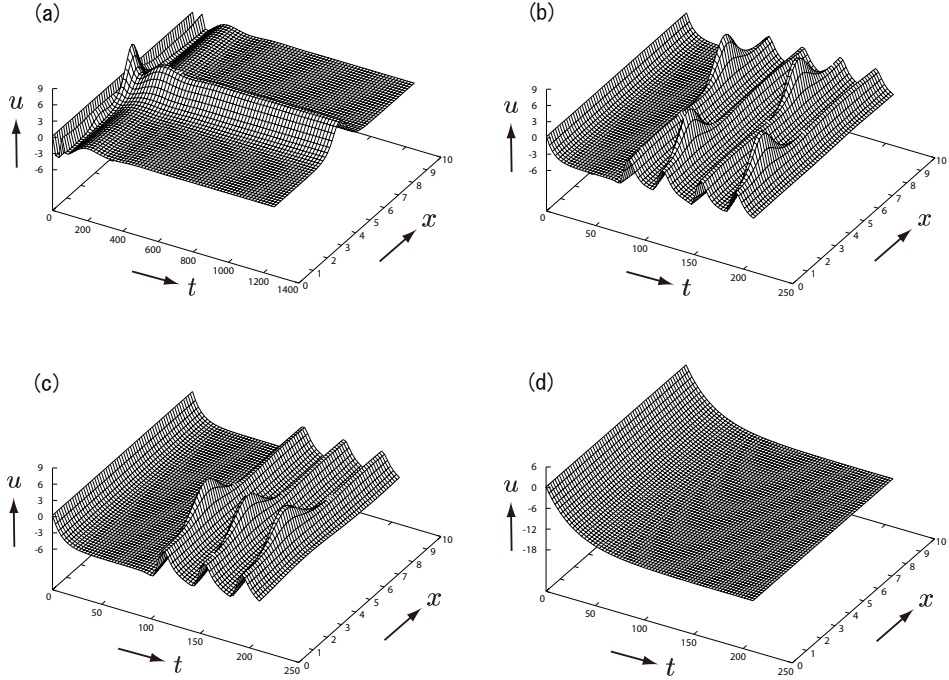}} 
\caption[]{ \label{zu3}
Spatiotemporal patterns appear in the second stage of the dynamics of numerical solutions for 
some values of $d_1$ when $\xi = 0.6$.
The initial value is given by $(\xi/2, \xi/2)$ with 
a small random perturbation. 
The values of $u(x,t)$ on $(x, t) \in [0, 10] \times [0, 1400]$ or
$(x, t) \in [0, 10] \times [0, 250]$
are represented by a 3D graph.
The profile of $v(x, t)$ is omitted here because the amplitude and spatial variation of $v(x, t)$ for each $t$
are relatively small as compared to those of $u(x, t)$. 
The numerical solution on $0 \leq x \leq 10$ presented in this figure can be regarded as
that on $-5 \leq x \leq 5$ under the periodic boundary condition. 
(a) $d_1 = 0.1$, A localized unimodal stationary pattern; 
(b) $d_1 = 0.25$, An oscillatory pattern that exhibits an alternating repetition of 
spatially homogeneous patterns and localized unimodal patterns with 
two different positions of their peak. The distance between two peak positions is (almost)
$K/2 = 5.0$. 
(c) $d_1 = 0.4$,  An oscillatory pattern that exhibits an alternating repetition of 
spatially homogeneous patterns and localized unimodal patterns with unique peak positions.
It should be noted that we cannot predict the position of the peak
of each localized unimodal pattern in (a)--(c)
because of a small random perturbation to the initial value of a solution.
(d) $d_1 = 0.8$, A spatially homogeneous stationary pattern.
}
\end{center}
\end{figure}

In the same way as the previous subsection, 
we perform numerical simulations for 
$d_1 = 0.1, \, 0.25, \, 0.4$ and $d_1= 0.8$ when $\xi = 0.6$.
Figure~\ref{zu3} shows that various spatiotemporal patterns appear
in the second stage of the dynamics of numerical solutions
when $\xi = 0.6$.
Figure~\ref{zu3} (a) shows a localized unimodal stationary pattern, which can be interpreted as 
the persistent maintenance of cell polarity.
Figures~\ref{zu3} (b) and (c) show 
stable limit cycles with large amplitude,
which exhibit an alternating repetition of unimodal and spatially homogeneous patterns. 
Figure~\ref{zu3} (c) shows that the periodically appearing
localized unimodal patterns 
take their maximum at $x = x_0$ for some $x_0 \in [-K/2, K/2]$ when $d_1 = 0.4$. 
In contrast, 
Figure~\ref{zu3} (b) shows that
the  periodically appearing localized unimodal patterns
take their maximum at $x = x_1$ or $x = x_2$ alternatively for some
$x_1 \in [-K/2, K/2]$ and $x_2 \in [-K/2, K/2]$ with
$| x_2 - x_1 | \approx K/2$ when $d_1 = 0.25$.
These limit cycles in Figures~\ref{zu3} (b) and (c) can be interpreted as cell polarity oscillations.
The limit cycle in Figure~\ref{zu3} (c) represents cell polarity oscillations 
with no reversal of polarity.
In contrast, noting that $x_2$ is antipodal to $x_1$,    
the limit cycle in Figure~\ref{zu3} (b) represents cell polarity oscillations
with the reversal of polarity.
We cannot predict the position of the peak
of each localized unimodal pattern in Figures~\ref{zu3} (a)--(c)
because of a small random perturbation to the initial value of a solution.
Figure~\ref{zu3} (d) shows a spatially homogeneous stationary pattern, which can be interpreted as nonpolarized cells.

\begin{remark}\label{rem2x} \rm
In the case of melanoma cells \cite[Figure 2]{PH}, 
spatially homogeneous stationary patterns as in Figure~\ref{zu3} (d) may be considered as
cells that randomly polarize into frequently changing directions.
\end{remark}

\begin{remark}\label{rem2xy} \rm
Recalling the fact that the ODE dynamics of \eqref{aaa1} with \eqref{aaa2}
does not qualitatively change when we vary $\alpha \in (0, 1)$, 
we can obtain similar spatiotemporal patterns as shown in Figure~\ref{zu3}
when we choose appropriate values of $d_1$ in \eqref{a1} 
with \eqref{a2} for fixed $\alpha$. 
The reason why we impose the condition \eqref{a2x} is 
that we can perform concrete calculations as seen in Subsections 3.1 and 5.1.
Moreover, our numerical computations of \eqref{a1} with \eqref{a2}
by using the pseudospectral method
are unlikely to cause numerical instabilities
under the condition \eqref{a2x} as compared to the case without \eqref{a2x}.
\end{remark}

\section{ Numerical bifurcation analysis}

In this section, we  numerically demonstrate 
that the bifurcation diagram of \eqref{a1} with \eqref{a2}
contains a structure that generates spatiotemporal patterns as presented in the last section.
By using AUTO~\cite{AUTO}, a software package used for studying the bifurcation structure of ODE systems, 
we investigate the bifurcation diagram of an ODE system, which is obtained by the finite Fourier series 
approximation for \eqref{a1} with \eqref{a2} in the same way as \cite{KI}. 
We investigate the bifurcation diagram of the ODE system with respect to $d_1$ for some fixed values of $\xi$ 
under the parameter values 
$a_1 = 0.5$, $a_2 = 2.2$ and $d_2 = 1.0$ by \eqref{b13}
and the interval length $K = 10.0$. \\

\begin{figure}[htbp]
\begin{center}
\includegraphics[width=70mm]{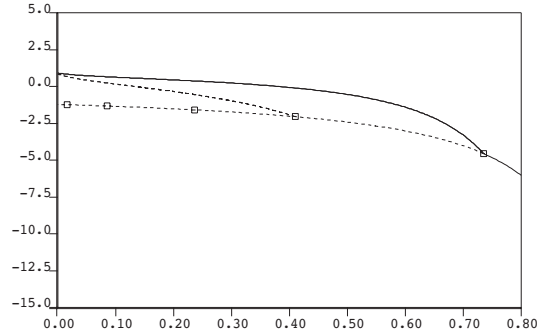}
\caption{Bifurcation diagram with respect to $d_1$ for $\xi=1.0$, where
$d_1$ decreases from $d_1 = 0.8$. 
The horizontal and vertical axes indicate $d_1$ and the size of the $u$-component of solutions represented by $\frac{1}{K}\int_I u(x)dx$, respectively. The solid line indicates stable solutions, whereas the dashed line indicates unstable ones. The white square indicates the pitchfork bifurcation point. Unstable branches bifurcating from
unstable branches for
$d_1 \leq 0.30$ are omitted.  
Notice that the two branches bifurcating from a pitchfork bifurcation point are piled up and are displayed in this bifurcation diagram.
The primary branch starting from $d_1 = 0.8$ represents a family of
spatially homogeneous equilibria \eqref{equx}, which correspond to 
spatially homogeneous stationary patterns as shown in Figure~\ref{zu3}(d).
The secondary branch bifurcating from the primary branch via the supercritical pitchfork bifurcation at $d_1 \approx 0.74$ represents
a family of (localized) unimodal stationary patterns as shown in Figure~\ref{zu3}(a). 
}
\label{100}
\end{center}
\end{figure}

Figure~\ref{100} shows the bifurcation diagram with respect to $d_1$ for $\xi=1.0$,
where $d_1$ decreases from $d_1 = 0.8$. 
The primary branch starting from $d_1 = 0.8$ represents a family of
spatially homogeneous equilibria 
\begin{equation}\label{equx}
(u(x),v(x))=\left( \frac{\xi^2 - a_2 }{(1- \alpha)\xi}, \frac{a_2 - \alpha\xi^2}{(1-\alpha)\xi} \right) 
\end{equation}
with $\alpha = d_1/d_2$, which is given by \eqref{aaaa4}.
By using a standard linear stability analysis, we can  
easily verify that these equilibria are stable for $d_1 > d_1^c$
whereas unstable for $0 < d_1 < d_1^c$, where $d_1^c$ is given by
\[
d_1^c = \frac{ a_1 a_2 d_2 K^2}{4\pi^2 + a_1 a_2 K^2}.
\]
It should be noted that $d_1^c$ is independent of $\xi$.
When $d_1 = 0.8$, we have $d_1^c \approx 0.74$, which 
gives a supercritical pitchfork bifurcation point detected by AUTO in Figure~\ref{100}.
We should consider that this bifurcation is due to a diffusion-driven instability
but not a well-known Turing instability. 
In fact, 
$d_1$, the diffusion coefficient of $u$, is also included in the nonlinear term $f(u, v)$
by $\alpha = d_1/d_2$.

\begin{remark}\label{rem2zz} \rm
When we consider \eqref{a1} with \eqref{a2} for fixed $\alpha$, we can similarly verify that a 
pitchfork bifurcation point where the stability of the equilibria 
\eqref{equx} changes is given by 
\[
\tilde{d}_1^c =  \frac{ a_1 d_2 (\alpha \xi^2 + a_2 ) K^2}{4\pi^2 +( a_1 \xi^2 + a_1 a_2 ) K^2}.
\]
In the strict sense, this bifurcation is not due to a Turing instability. In fact, as seen 
in Section 2, 
the equilibria \eqref{equx} are stable in the ODE \eqref{aaa1} obtained from \eqref{a1} by dropping the diffusion terms. However, the 
linearized operator of the right hand side of 
\eqref{aaa1} at \eqref{equx} has zero eigenvalue, which implies
that a well-known condition for Turing instability is not satisfied (cf. \cite{Mu}).
Therefore, the bifurcation in this case is sometimes called a Turing-like bifurcation \cite{Is}.
\end{remark}

The secondary branch bifurcates from the primary branch via the supercritical pitchfork bifurcation at $d_1 \approx 0.74$
as $d_1$ decreases. This stable branch represents a family of (localized) unimodal 
stationary patterns as shown in Figure~\ref{zu3}(a). 
The bifurcation diagram presented in Figure~\ref{100} suggests that \eqref{a1} with \eqref{a2} has no limit cycles
as shown in Figures~\ref{zu3}(b) and (c). 
This type of bifurcation diagram can be obtained for $\xi \gtrsim 0.67$.

\begin{figure}[htbp]
\begin{center}
\includegraphics[width=70mm]{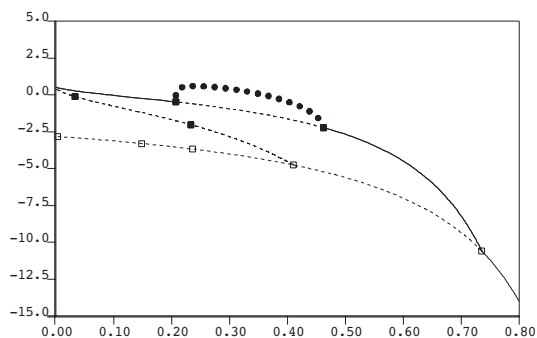}
\caption{Bifurcation diagram with respect to $d_1$ for $\xi=0.64$, where
$d_1$ decreases from $d_1 = 0.8$. 
This bifurcation diagram is presented in the same way as that in Figure~\ref{100}.
The horizontal and vertical axes indicate $d_1$ and the size of the $u$-component of solutions represented by
$\frac{1}{K}\max_{t>0} \int_I u(x,t)dx$, respectively. 
The white and black squares indicate the pitchfork and Hopf bifurcation points, respectively. 
The curve represented by a family of the black circles $\bullet$ indicates the stable branch of periodic solutions.
The primary branch starting from $d_1 = 0.8$ represents a family of
spatially homogeneous equilibria \eqref{equx}, which correspond to 
spatially homogeneous stationary patterns as shown in Figure~\ref{zu3}(d).
The secondary branch bifurcating from the primary branch via the supercritical pitchfork bifurcation at $d_1 \approx 0.74$ represents
a family of (localized) unimodal stationary patterns as shown in Figure~\ref{zu3}(a). 
The third branch (indicated by $\bullet$) bifurcating from the second branch via the Hopf bifurcations at $d_1 \approx 0.46$ and $d_1 \approx 0.21$ represents
a family of limit cycles as shown in Figure~\ref{zu3}(c). 
 }
\label{064}
\end{center}
\end{figure}

Figure~\ref{064} shows the bifurcation diagram for $\xi=0.64$. 
As seen in Figure~\ref{100}, the secondary branch consisting of (localized) unimodal stationary patterns
bifurcates from the primary branch consisting of spatially homogeneous equilibria 
via a supercritical pitchfork bifurcation at $d_1 \approx 0.74$ as $d_1$ decreases.
However, (localized) unimodal stationary patterns on 
the second branch are unstable for $0.21 \lesssim d_1 \lesssim 0.46$.
In fact, the third branch consisting of stable limit cycles as shown in Figure~\ref{zu3}(c)
bifurcates from the secondary branch via Hopf bifurcations at $d_1 \approx 0.46$ and $d_1 \approx 0.21$.
In this case, we cannot observe limit cycles as shown in Figure~\ref{zu3}(b). 
This type of bifurcation diagram can be obtained for $ 0.61 \lesssim \xi \lesssim 0.66$.

\begin{figure}[htbp]
\begin{center}
\includegraphics[width=70mm]{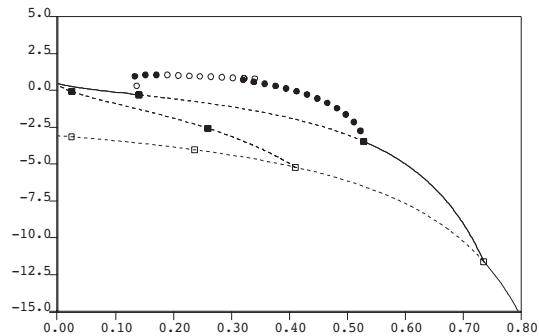}
\caption{Bifurcation diagram with respect to $d_1$ for $\xi=0.60$, where
$d_1$ decreases from $d_1 = 0.8$.  
This bifurcation diagram is presented in the same way as that in Figure~\ref{064}. 
The curve represented by a family of the black circles $\bullet$ indicates the stable branch of periodic solutions,
whereas that of the white circles $\circ$ indicates the unstable one.
The primary branch starting from $d_1 = 0.8$ represents a family of
spatially homogeneous equilibria \eqref{equx}, which correspond to 
spatially homogeneous stationary patterns as shown in Figure~\ref{zu3}(d).
The secondary branch bifurcating from the primary branch via the supercritical pitchfork bifurcation at $d_1 \approx 0.74$ represents
a family of (localized) unimodal stationary patterns as shown in Figure~\ref{zu3}(a). 
The third branch (indicated by $\bullet$ and $\circ$) bifurcating from the second branch via the Hopf bifurcations at $d_1 \approx 0.53$ and $d_1 \approx 0.14$ represents
a family of limit cycles as shown in Figure~\ref{zu3}(c). 
At present, we cannot trace the fourth branch bifurcating from the third branch via the period doubling bifurcations at $d_1 \approx 0.31$ and $d_1 \approx 0.18$. This branch that is not displayed here corresponds to 
a family of limit cycles as shown in Figure~\ref{zu3}(b). 
 }
\label{060}
\end{center}
\end{figure}

Figure~\ref{060} shows the bifurcation diagram for $\xi=0.60$.
As seen in Figure~\ref{064}, the third branch consisting of limit cycles as shown in Figure~\ref{zu3}(c)
bifurcates from the second branch consisting of (localized) unimodal stationary patterns
via Hopf bifurcations at $d_1 \approx 0.53$ and $d_1 \approx 0.14$.
However, the third branch has unstable limit cycles for $0.13 \lesssim d_1 \lesssim 0.14$
and $0.18 \lesssim d_1 \lesssim 0.31$. 
We can observe limit cycles as shown in Figure~\ref{zu3}(b) for $0.18 \lesssim d_1 \lesssim 0.31$, and expect that
the fourth branch consisting of these limit cycles bifurcates from the third branch at $d_1 \approx 0.18$ and
$d_1 \approx 0.31$ via period doubling bifurcations. In fact, 
AUTO detects a Floquet multiplier passing the unit circle $|\lambda| = 1$
through $\lambda = -1$ around $d_1 \approx 0.18$ and
$d_1 \approx 0.31$. At present, it is technically difficult to trace this fourth branch by using AUTO. 
We can obtain similar bifurcation diagram as shown in Figure~\ref{060} for $\xi \lesssim 0.60$.

\begin{figure}[htbp]
\begin{center}
\begin{tabular}{cc}
\includegraphics[width=60mm]{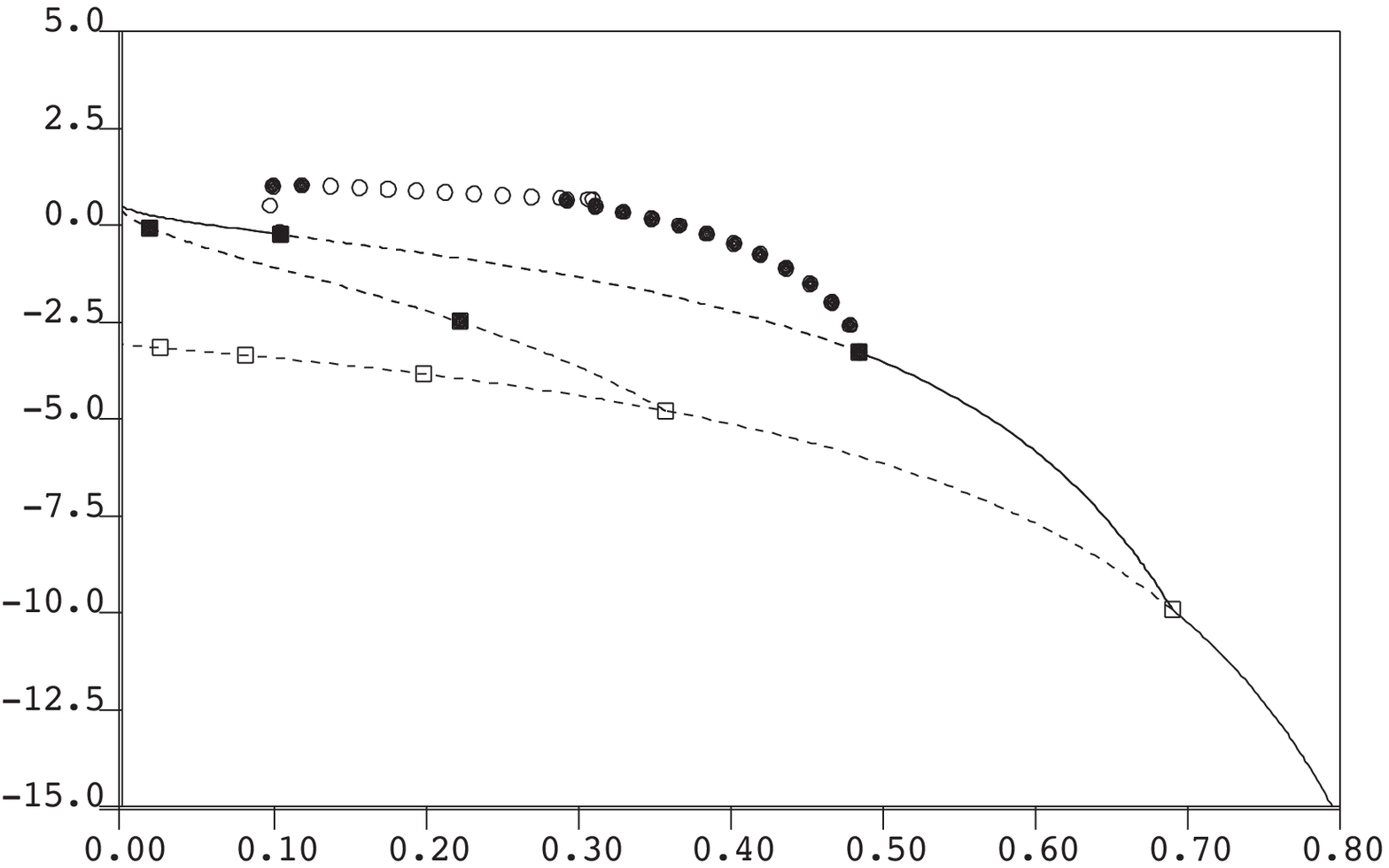} & 
\includegraphics[width=60mm]{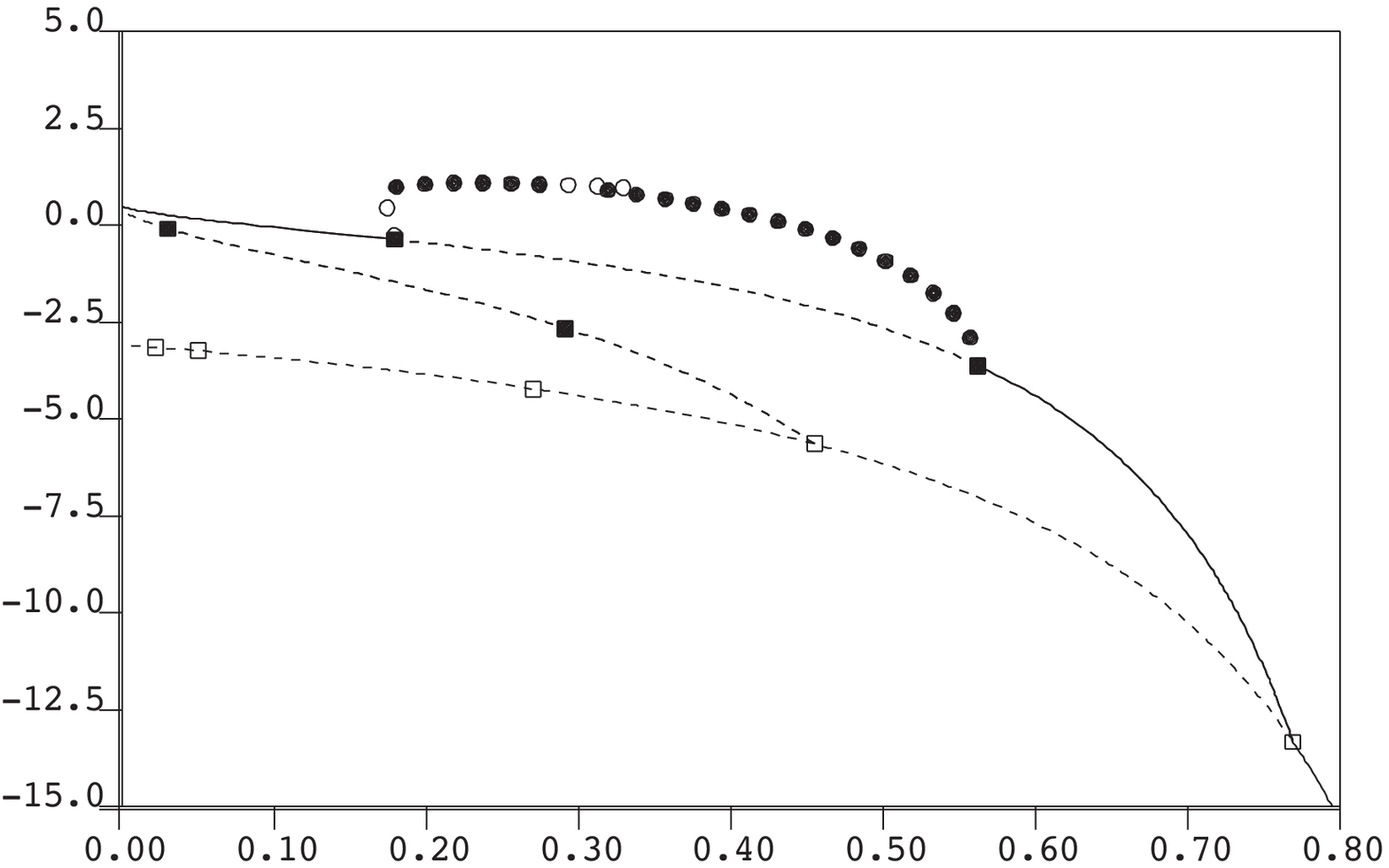}\\
(a) \ $a_1=0.4$ & (b) \ $a_1=0.6$\\
\includegraphics[width=60mm]{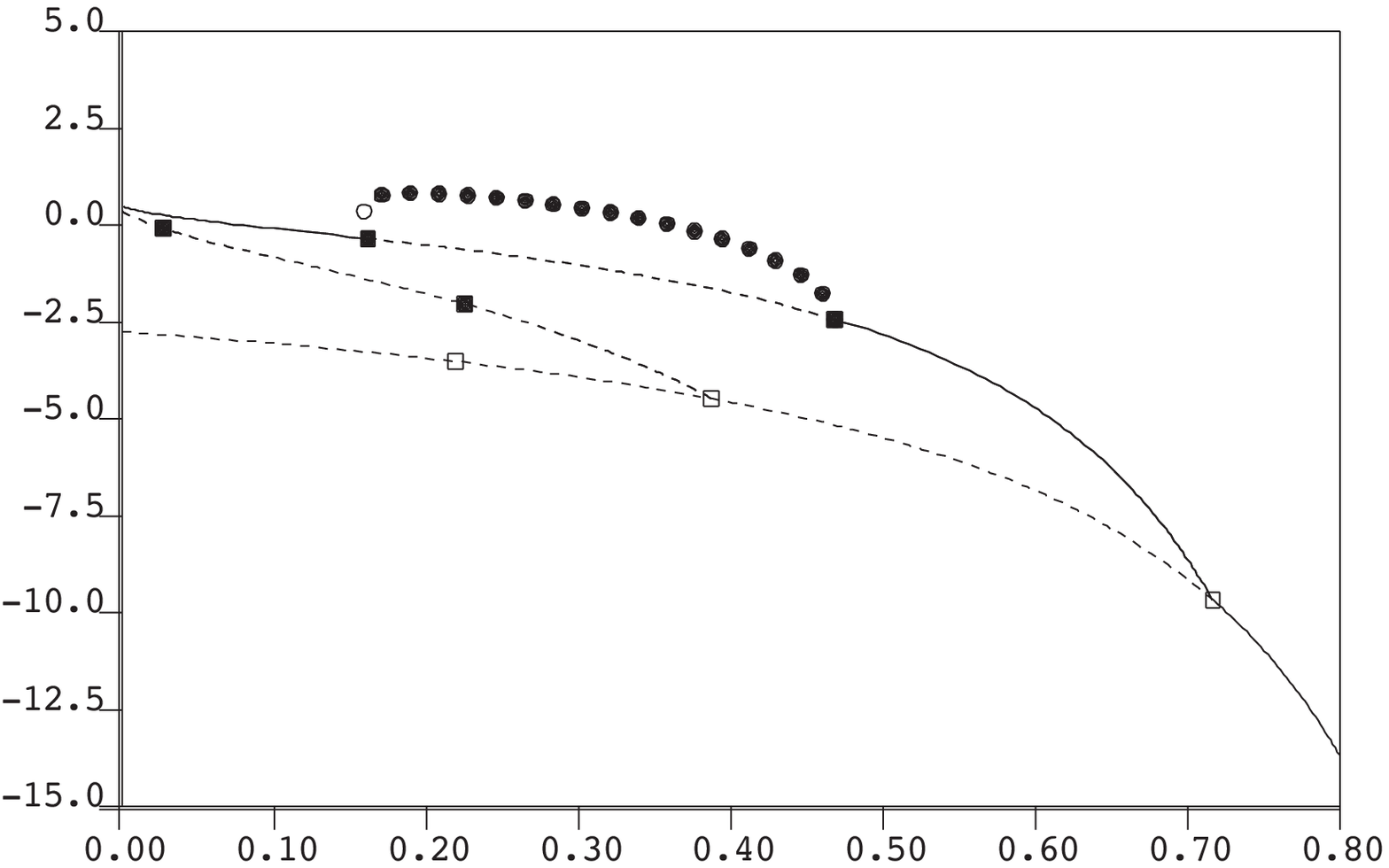} & 
\includegraphics[width=60mm]{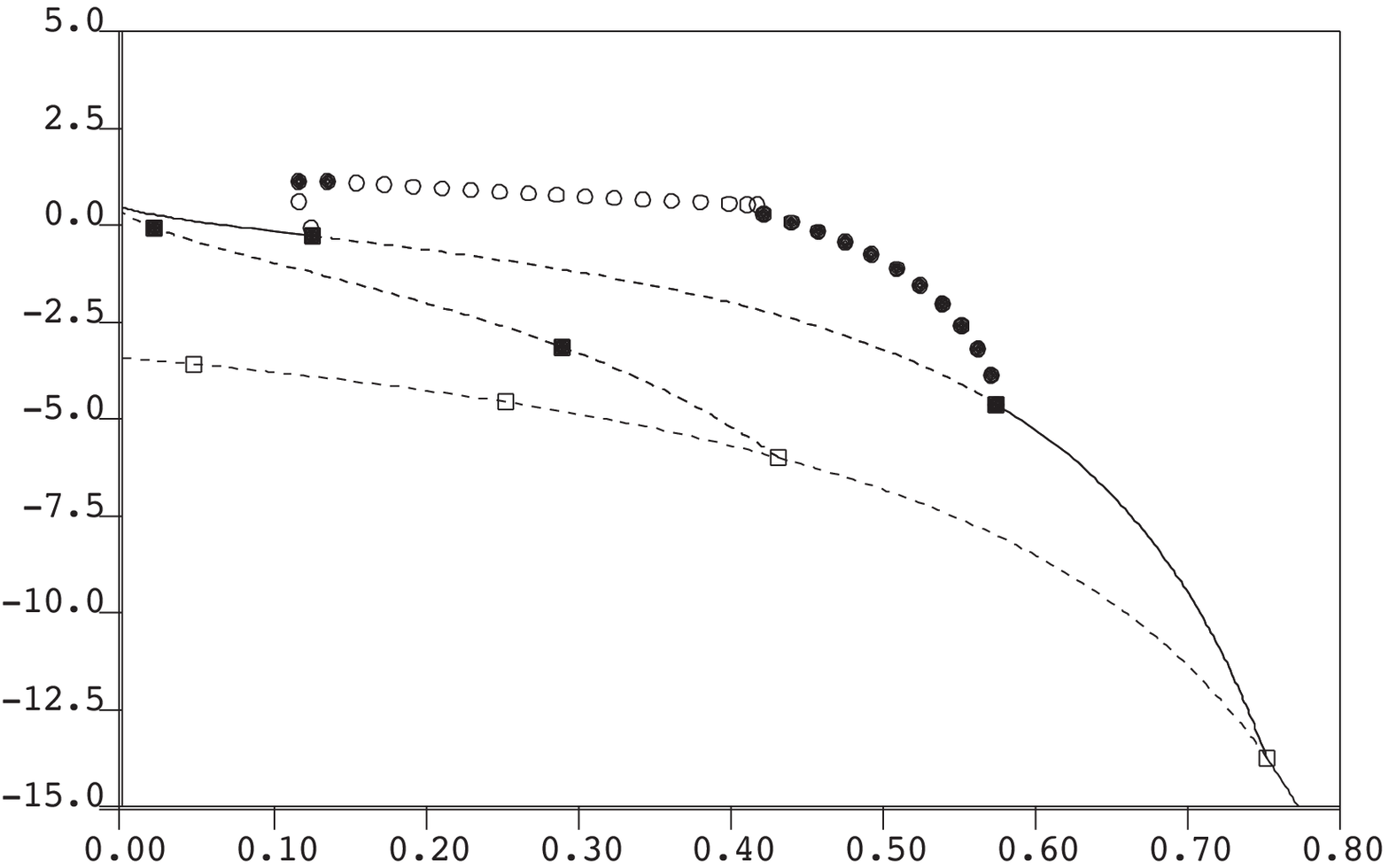}\\
(c) \ $a_2=2.0$ & (d) \ $a_2=2.4$\\
\includegraphics[width=60mm]{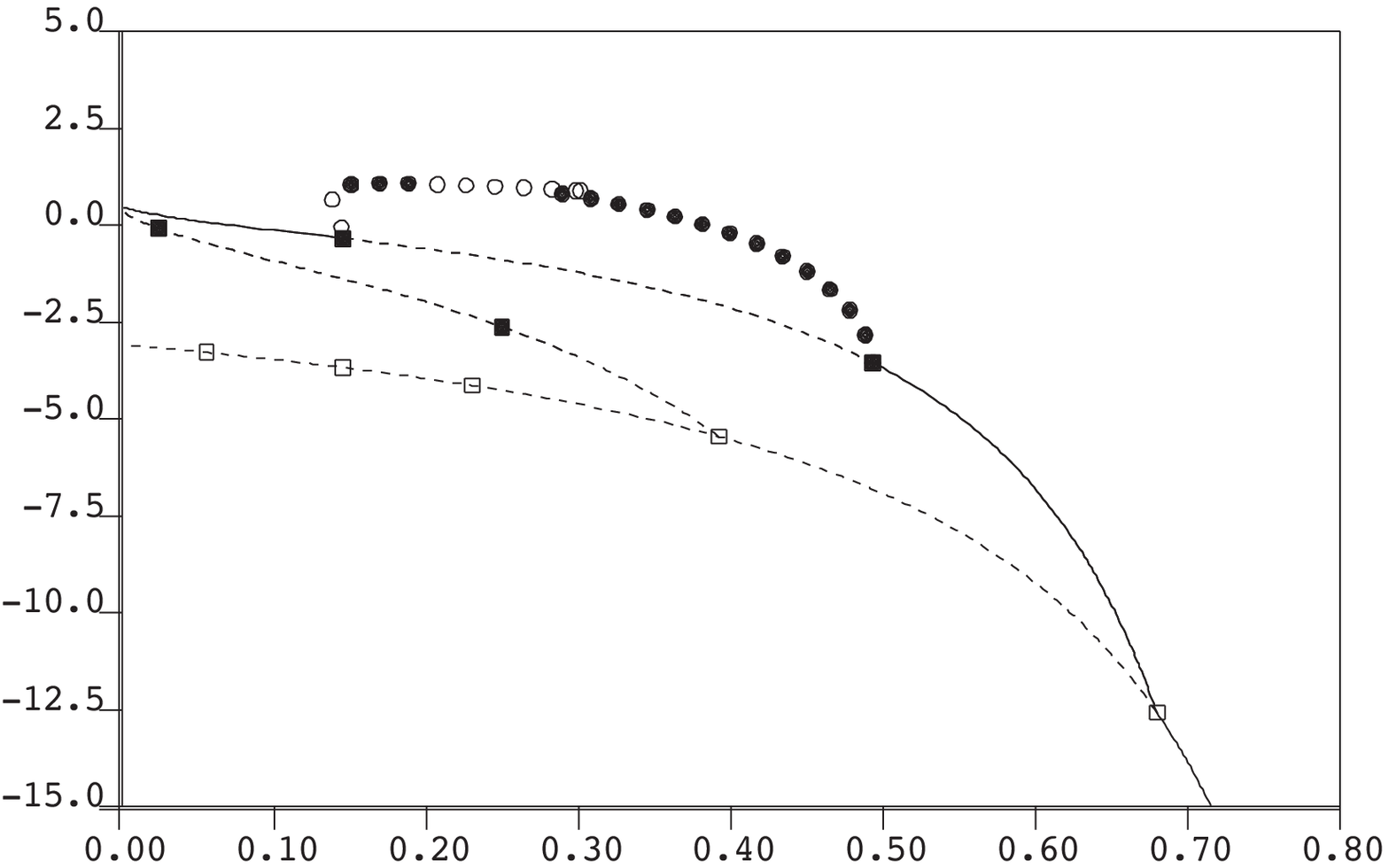} & 
\includegraphics[width=60mm]{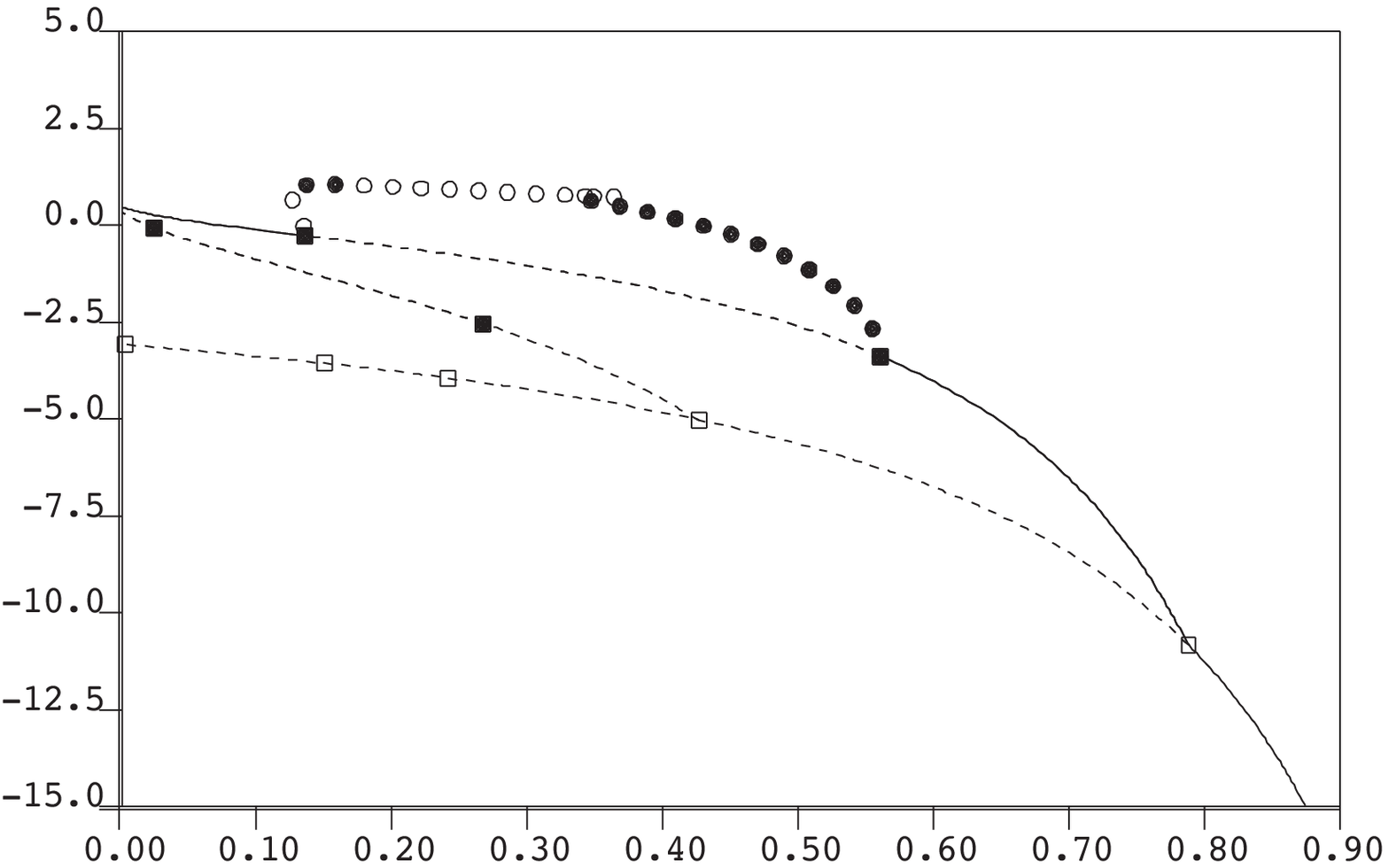}\\
(e) \ $d_2=0.9$ & (f) \ $d_2=1.1$
\end{tabular}
\caption{Bifurcation diagrams with respect to $d_1$ for $\xi=0.60$ and 
the various values of $a_1$, $a_2$ and $d_2$. 
These bifurcation diagrams are presented in the same way as that in Figure~\ref{060}.
The parameter values except for the indicated value at the bottom of each figure are the same as those used in the bifurcation diagram presented in Figure~\ref{060},
where $a_1 = 0.5$, $a_2 = 2.2$ and $d_2 = 1.0$ by \eqref{b13}. 
The bifurcation diagrams in (a) and (b) show that the bifurcation diagram in Figure~\ref{060}
does not qualitatively change when we vary the value of $a_1$ near $a_1 = 0.5$. 
The bifurcation diagrams in (c) and (d) show that it
does not qualitatively change when we vary the value of $a_2$ near $a_2 = 2.2$. 
The bifurcation diagrams in (e) and (f) show that it
does not qualitatively change when we vary the value of $d_2$ near $d_2 = 1.0$. 
}
\label{others}
\end{center}
\end{figure}

We can obtain bifurcation structures similar to those 
as described above near the specific parameter values \eqref{b13}.  
Figure~\ref{others} shows 
that the bifurcation diagram with respect to $d_1$ for $\xi=0.60$ 
does not qualitatively change when we vary
one of the values $a_1$, $a_2$, and $d_2$ near the specific values.
Similarly, we can obtain the same results concerning the bifurcation
diagrams with respect to $d_1$ for other values of $\xi$.   
Therefore, we conclude that the bifurcation structures presented in this section are 
structurally stable.

\section{Dynamics of spatiotemporal patterns of perturbed system}

In this section, we investigate the dynamics of spatiotemporal patterns 
in the perturbed system \eqref{a4} with \eqref{a2} and \eqref{a5}.
Our results are useful for understanding the effects of extracellular signals on
cell polarization.

\subsection{ Motion of localized unimodal pattern}

\begin{figure}
\begin{center}
\resizebox{13cm}{5cm}{\includegraphics{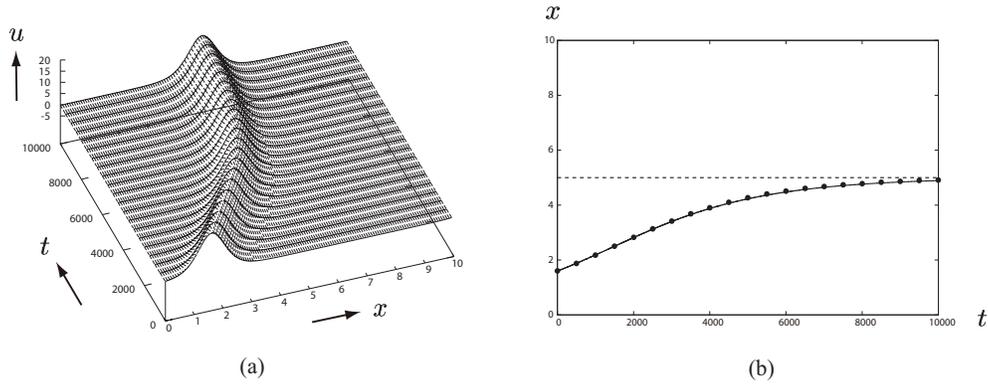}} 
\caption[]{ \label{zu5}
The motion of a localized unimodal pattern in the perturbed system \eqref{a4A}.
The initial value is given by $(\bar{u}(x-c), \bar{v}(x-c))$ with $c=1.6$,
where $\bar{u}$ and $\bar{v}$ are given by \eqref{b9} and \eqref{b10}, respectively.
(a) The dynamics of $u$-component.
The values of $u(x,t)$ on $0 \leq x \leq 10$ and $0 \leq t \leq 10000$ are 
represented by a 3D graph. 
The profile of $v(x, t)$ is omitted here because the amplitude and spatial variation of $v(x, t)$ for each $t$
are relatively small as compared to those of $u(x, t)$. 
(b) The comparison between the theory and numerical simulations.
The rigid and dashed lines represent a solution of \eqref{c6} with the initial value $\ell(0) = 1.6$
and its limit given by $\ell = 5.0$, respectively. 
The dots represent the maximal point of $u(x, t)$ for $t = 500 n \ (n \in \text{\bf Z}, 0 \leq n \leq 20)$. 
}
\end{center}
\end{figure}

In the same way as Subsection 3.2, we find that
a solution of \eqref{a4} starting from 
a spatially homogeneous initial state with a small disturbance 
quickly and temporally approaches 
$(\bar{u}(x-c), \bar{v}(x-c))$ for some $c \in [-K/2, K/2)$
under certain conditions, where $\bar{u}$ and $\bar{v}$ are given by \eqref{b9}
and \eqref{b10}, respectively. Then, the solution
translationally moves to the maximum point of $a_E(x)$, as shown in Figure~\ref{zu5}.   
We derive the equation of motion for this translational movement 
according to \cite{EKS}. In this subsection, to avoid confusion, we rewrite \eqref{a1} and \eqref{a4}
as 
\begin{equation}\label{a1A}
\left \{
\begin{array}{rcl}
\dot{u} & = & d_1 u_{xx} -  f(u,v)  \\[1ex]
\dot{v} & = & d_2 v_{xx}  +  f(u,v)
\end{array} 
\right.
\end{equation}
and
\begin{equation}\label{a4A}
\left \{
\begin{array}{rcl}
\dot{u} & = & d_1 u_{xx} - \{  f(u,v) - \ep g(x, u, v) \} \\[1ex]
\dot{v} & = & d_2 v_{xx}  + \{ f(u,v) - \ep g(x, u, v)  \},
\end{array} 
\right.
\end{equation}
respectively, and \eqref{a2} as  
\begin{equation}\label{a2A}
f(u, v) = - a_1 ( u + v) \{ (\alpha u + v) (u + v) - a_2 \}.
\end{equation}
Moreover, to obtain a concrete and exact expression of
the equation of motion, we consider that
\begin{equation}\label{a5A}
g(x, u, v) = a_1 (u+v) a_E(x),
\end{equation}
where
\begin{equation}\label{c0}
a_E(x) = A \cos ( 2\pi x /K)
\end{equation}
takes its maximum at $x=0$.
In this subsection, we carry our
argument based on \cite{EKS} by using \eqref{a1A}--\eqref{c0}.

We consider a solution that is well approximated by $S(x - \ell(t) )$, where 
\begin{equation}\label{c1}
S(x) := ( \bar{u}(x), \bar{v}(x) ) = (\varphi_1(x), \varphi_2(x) ) + O(e^{-bK}), 
\end{equation}
and $\varphi_1(x)$ and $\varphi_2(x)$ are given by \eqref{b8} and \eqref{b8x}, respectively.
Hereafter, we derive the equation for $\ell(t)$ by applying an argument in \cite{EKS}.

\begin{remark}\label{rem2Z} \rm
The theory of \cite{EKS} was built on a mathematical assumption \cite[Assumption 3.4]{EKS},
which guarantees that solutions stay in a small neighborhood of a manifold
$M = \{ \, S(x-\ell) \, | \, \ell \in \text{\bf R} \, \}$.
Although such an assumption is indispensable to build a rigorous theory (cf. \cite{Hen}), 
it is technically difficult
to check it in many practical problems. Here, we formally apply an argument in 
\cite{EKS} to our problem without checking the mathematical assumption. 
\end{remark}

First, we consider conditions that
determine the sign of $\lg S_z, \Phi^* \rg$ to be defined by \eqref{c5Y}.
It follows from \eqref{c1} that
\[
S_x(x) = w(x) ( d_2, -d_1), 
\]
where 
\begin{equation}\label{c3}
w(x) = -\frac{2 b p_0}{ d_2 - d_1} \text{sech}^2(bx) \text{tanh}(bx) + O(e^{-bK})
\end{equation}
with
\begin{equation}\label{b7yy}
b = \dis\frac{1}{2} \sqrt{ \dis\frac{(d_2 - d_1)a_1 a_2}{d_1 d_2}   }
\ \ \ \ \text{and} \ \ \ \  
p_0 = \dis\frac{bK}{2} \xi + O(Ke^{-bK})
\end{equation}
by \eqref{b7} and \eqref{b12xx}.
Therefore, we see that $w(x) <0$ for $x >0$ as mentioned in \cite[Lemma 3.2]{EKS}.

Since $f_v(u, v) = -2a_1(u+v)(\alpha u+v) -a_1(u+v)^2 + a_1 a_2$ by \eqref{a2A}, 
noting \eqref{b11}, \eqref{b11xx} and $p_0 q_0 = 3a_2/2$ by \eqref{b7}, we have
\[
\begin{array}{rcl}
h_2(x) & := & -\dis\frac{f_v( \bar{u}(x), \bar{v}(x)) w(x)}{d_2}  
\\[2ex]
& = & \dis\frac{ 2bp_0 }{ d_2(d_2-d_1)} \text{sech}^2(bx) \text{tanh}(bx)
F(x) + O(Ke^{-bK})
\end{array}
\]
where $F(x) := -3a_1a_2 \text{sech}^2(bx) - a_1 p_0^2 \text{sech}^4(bx) + a_1 a_2 $.
Let  $t = \text{sech}^2(bx)$. Then, we have $F(x) = g(t)$, where
$$
g(t) := -3a_1 a_2 t - a_1 p_0^2 t^2 + a_1 a_2.
$$
We see that $F(x)$ is monotone increasing in $x$ for $x>0$ because
$t = \text{sech}^2(bx)$ is monotone decreasing in $x$ for $x >0$ and 
$g(t)$ is monotone decreasing in $t$ for $t >0$.
Moreover, we have 
$F(0) = g(1) = -2a_1 a_2 - a_1 p_0^2 <0$ and
$F(K/2) = g(0) + O(e^{-bK}) = a_1 a_2 + O(e^{-bK}) >0$.
Therefore, there exists $\beta \in (0, K/2)$ such that $h_2(x) < 0$ for $0 < x< \beta$
and $h_2(x) > 0$ for $\beta <x < K/2$. 
Furthermore, noting \eqref{b7yy}, a direct calculation 
with the aid of Mathematica shows 
$$
\begin{array}{l}
\dis\int_0^{K/2} x h_2(x) dx =
 \dis\frac{2 b a_1 p_0 }{d_2 (d_2 -d_1)} \Big{\{}
-3a_2 \dis\int_0^{\infty} x \, \text{sech}^4(bx)\text{tanh}(bx) dx 
\\[3ex]
\ \ \  -  p_0^2 \dis\int_0^{\infty} x \,  \text{sech}^6(bx)\text{tanh}(bx) dx 
+ a_2 \dis\int_0^{\infty} x \, \text{sech}^2(bx)\text{tanh}(bx) dx
\Big{\}} 
\\[3ex]
\ \ \  \ \ \ \ \ \ \ \ \ \ + O(Ke^{-bK})
\\[3ex]
= 
 \dis\frac{2 b a_1 p_0}{d_2 (d_2 -d_1)} \Big{\{}
 -3a_2 \cdot \frac{1}{6b^2} -  p^2_0 \cdot \frac{4}{45b^2} + a_2 \cdot \frac{1}{2b^2} \Big{\}}
 + O(Ke^{-bK})
\\[3ex]
= -\dis\frac{8 a_1 p_0^3}{45 (d_2 -d_1) d_2 b  }  + O(Ke^{-bK}) 
 = 
 -\dis\frac{4 a_1 p_0^2 K \xi}{45 (d_2 -d_1) d_2   }  + O(Ke^{-bK}).
\end{array}
$$
Therefore, 
$$
\dis\int_0^{K/2} x h_2(x) dx < 0 
$$
holds if
$$
\xi = \dis\frac{1}{K} \int_I ( u(x, 0) + v(x, 0) ) dx >0 .
$$
Hence, the condition (ii) of \cite[Proposition 3.6]{EKS} holds. Thus, we see 
that $\varphi_2^*(x) > 0 $ for $0 < x < K/2$, where 
$\varphi_2^*$ is an odd periodic function with the period $K$, which satisfies
$(\varphi_2^*)'' = h_2$,
$\varphi_2^*(0) = \varphi_2^*(K/2) = 0$, and
\begin{equation}\label{c550}
\frac{d\varphi_2^*}{dx}(K/2) = \dis\frac{2}{K} \int_0^{K/2} x h_2(x) dx.
\end{equation}

Next, we derive the equation for $\ell(t)$. 
It follows from \cite[Equation (4.3)]{EKS} that 
\begin{equation}\label{c5}
\frac{d \ell}{dt } =- \ep \frac{ J(\ell)}{ \lg S_z, \Phi^* \rg } + O(\ep^2),
\end{equation}
where 
\begin{equation}\label{c5X}
J(\ell) :=  - \dis\int_0^{K/2} a_1 ( \bar{u} (z) + \bar{v}(z) ) w(z) \{ a_E( z+ \ell) - a_E( z -\ell) \} dz
\end{equation}
and 
\begin{equation}\label{c5Y}
\lg S_z, \Phi^* \rg := -2d_2 \dis\int_0^{K/2} w^2(z) dz + 2(d_2 -d_1) \dis\int_0^{K/2} w(z) \varphi_2^*(z) dz.
\end{equation}
The calculations of the right hand sides of \eqref{c5X} and \eqref{c5Y}
are rather lengthy, and found in the appendix.  
As a result, noting the second equation of \eqref{b7yy}, it follows from \eqref{c5} that 
\begin{equation}\label{c5Z}
\frac{d \ell}{dt } =- \ep \left( C \sin ( \frac{2 \pi \ell}{K} ) + O(Ke^{-bK}) \right) + O(\ep^2),
\end{equation}
where $C$ is given by
$$
C = \dis\frac{ 5 a_1 d_2 \pi^3 A ( b^2 K^2 + \pi^2) \,  \text{cosech} (\dis\frac{\pi^2}{bK} )}
{ b^3 K^4 \Big{\{} \dis\frac{4 b^2 d_2^2}{d_2 - d_1} 
+ a_1 a_2 + \dis\frac{2 a_1 bK \xi^2}{21} (3bK -7) \Big{\}}  }
$$
with the first equation of \eqref{b7yy}.
For example, the value of $C$ is given by $C \approx 6.35 \times 10^{-3}$ under the 
same condition in the simulation by \cite{Ot}, i.e., 
$K=10.0$, $\xi = 2.0$, $d_1 = 0.1$, $A = 1.1$ and the parameter values \eqref{b13}. 

We now examine the validity of \eqref{c5Z} from a viewpoint of numerical simulations.
In a similar manner to the previous section, we numerically solve \eqref{a4A} 
with $a_E(x) = A\cos ( \frac{2\pi }{K} ( x - \frac{K}{2} ) )$ on $0 \leq x \leq K$ under the periodic boundary condition.
In this case, neglecting error terms, \eqref{c5Z} is replaced by
\begin{equation}\label{c6}
\frac{d \ell}{dt } = - \ep C \sin \left( \frac{2 \pi }{K}( \ell - \frac{K}{2})  \right).
\end{equation}
Figure~\ref{zu5} shows that a localized unimodal pattern moves to the maximal point of $a_E(x)$ following the equation of 
motion given by \eqref{c6}. 
Here, the values of parameters are given by \eqref{b13}, and
$K=10.0$, $\xi = 2.0$, $d_1 = 0.1$, $A = 1.1$ and $\ep = 0.1$.
The initial value is given by $(\bar{u}(x-c), \bar{v}(x-c))$ with $c=1.6$,
where $\bar{u}$ and $\bar{v}$ are given by \eqref{b9} and \eqref{b10}, respectively.
We note that these results are valid for $0 < \ep  \lesssim  0.1$. 
This suggests that the translational movement of cell polarity can be controlled by sufficiently weak extracellular signals.

\begin{remark} \rm
\cite{Ot} studied \eqref{a4A} by setting
$a_2 \to a_2 \left( 1 + \frac{\ep}{2} \sin ( \frac{2\pi x}{K} ) \right)$ in \eqref{a2A},
which is equivalent to $a_E(x) = A \cos ( \frac{2\pi }{K} ( x + \frac{K}{4} ) )$ with 
$A = a_2/2$. Their formal analysis for the translational movement of a localized unimodal 
pattern
could not give a precise result because their analysis was 
done under the assumption in which the speed of the movement is constant. 
\end{remark}

\subsection{ Dynamics of spatiotemporal oscillatory patterns}

The result in the last subsection suggests that 
the position of the peak of the final localized unimodal stationary
pattern of \eqref{a4} can be determined by the maximum point of $a_E(x)$.
In this subsection, we investigate whether the perturbed system \eqref{a4} has
spatiotemporal oscillatory patterns as shown in Figures~\ref{zu3} (b) and (c)
under the same conditions imposed on the unperturbed system \eqref{a1}.
In particular, we examine whether
the position of the peak of such periodically appearing unimodal pattern
can be determined by the maximum point of $a_E(x)$.

We consider \eqref{a4} 
with \eqref{a5} and 
$a_E(x) = A\cos ( \frac{2\pi }{K} ( x - \frac{K}{2} ) )$ on $0 \leq x \leq K$ under the periodic boundary condition.
In the same way as Subsection 3.3, 
we numerically solve \eqref{a4} for 
$d_1 = 0.1, \, 0.25, \, 0.4$ and $d_1= 0.8$ when $\xi = 0.6$
under the parameter values \eqref{b13} and 
$K=10.0$,  $A = 1.1$ and $\ep = 0.1$.

\begin{figure}
\begin{center}
\resizebox{125mm}{90mm}{\includegraphics{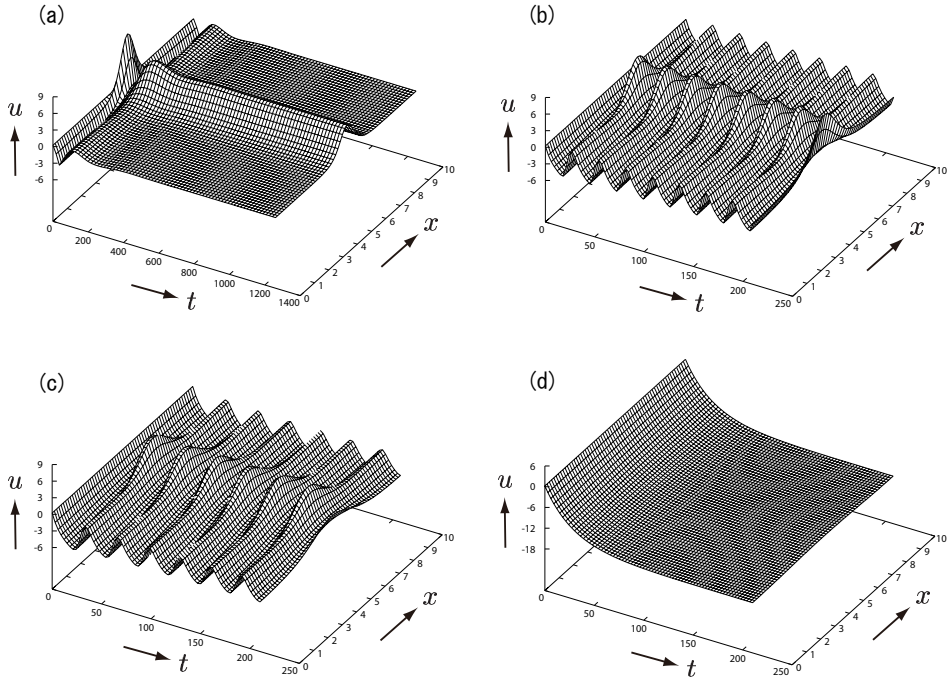}} 
\caption[]{ \label{zu4}
Spatiotemporal patterns appear for some values of $d_1$ when $\xi = 0.6$
and $\ep = 0.1$.
The initial value is given by $(\xi/2, \xi/2)$ with a small random perturbation. 
The 3D graphs are presented in the same way as those in Figure~\ref{zu3}.
(a) A localized unimodal stationary pattern for $d_1 = 0.1$; 
(b), (c) An oscillatory pattern that exhibits an alternating repetition of spatially homogeneous patterns and localized unimodal patterns with unique 
peak positions. The values of $d_1$ are given by $d_1 = 0.25$ and $d_1 = 0.4$, respectively.
In contrast to the case of Figure~\ref{zu3}, 
it should be noted that the position of the peak
of each localized unimodal pattern in (a)--(c) 
is determined by the maximum point of $a_E(x)$
without depending on a small random perturbation to the initial value of a solution.
(d) A spatially homogeneous stationary pattern for $d_1 = 0.8$.
}
\end{center}
\end{figure}

Figure~\ref{zu4} shows that the perturbed system \eqref{a4} exhibits 
similar spatiotemporal patterns as shown in Figure~\ref{zu3}.
In contrast to the case of the unperturbed system \eqref{a1},
the position of the peak of each localized unimodal pattern of the perturbed system \eqref{a4} 
is determined by the maximum point of $a_E(x)$, i.e.,
it does not depend on a small random perturbation to the initial value of a solution. 
Moreover, we note that the stable limit cycles in the perturbed system \eqref{a4} 
earlier appear and their period is shorter as compared to those in the unperturbed system \eqref{a1}.

When $d_1=0.4$, we can observe oscillatory spatiotemporal patterns as shown in Figure~\ref{zu4} (c) for $0 < \ep \leq 0.1$.
In contrast, when $d_1 = 0.25$, we cannot always observe oscillatory spatiotemporal patterns
as shown in Figure~\ref{zu4} (b) for $0 < \ep \leq 0.1$. More precisely, 
we can observe them for $0.015 \lesssim \ep \leq 0.1$, i.e., the position of 
the peak of each localized unimodal pattern
cannot be determined by the maximum point of $a_E(x)$ for $0 < \ep \lesssim 0.014$. 
These results suggest that 
cell polarity oscillations 
can be controlled by sufficiently weak extracellular signals
if the reversal of polarity does not occur.

\section{Discussion}

In this paper, we show that a reaction-diffusion system with bistable nonlinearity 
and mass conservation, proposed by \cite{Ot}, 
exhibits four different spatiotemporal patterns
including two types of oscillatory patterns, which can be regarded as 
cell polarity oscillations with the reversal and non-reversal of polarity, respectively.
The trigger causing these patterns is a diffusion-driven (Turing-like) instability. 
Moreover, we show that the periodic reversal is due to a period doubling bifurcation 
using a numerical bifurcation analysis. 
Furthermore, we investigate the effects of extracellular signals on 
cell polarity oscillations.
Our results suggest that cell polarity oscillations 
can be controlled by sufficiently weak extracellular signals
if the reversal of polarity does not occur.

To understand the biological implications of our results based on this model,
we consider a cell (simplified and abstract model) with the following properties:
(1) Cell polarity is controlled by chemical $X$.
(2) The amount of $X$ is conserved.
(3) Chemical $X$ has two states $X_m$ and $X_c$.
(4) The diffusion coefficient of $X_c$ is faster than that of $X_m$.
(5) The cell size $K$ is conserved.
Let $u = [X_m] - c_m$ and $v = [X_c] - c_c$, where
$[X_m]$ and $[X_c]$ denote the concentrations of $X_m$ and $X_c$, respectively,
and $c_m$ and $c_c$ are the reference values of $[X_m]$ and $[X_c]$,
respectively. Then, noting the conservation law \eqref{ax2}, we have
$$
\begin{array}{l}
\xi = \dis\frac{1}{K}\int_0^K (u + v) dx = |X|/K - C,
\end{array}
$$
where $|X|$ denotes the amount of $X$, and $C = c_m + c_c $ is a positive constant independent of
$|X|$ and $K$. 
Recalling that we consider $\xi = |X|/K - C > 0 $ as a parameter in our numerical 
simulations and bifurcation analysis,
our results are then summarized as follows:
(1) Cell polarity can be established for the appropriate values of the diffusion coefficients of 
$X_m$ and $X_c$.
(2) Cell polarity oscillations can be observed for the appropriate values of the diffusion coefficients of $X_m$ and $X_c$ when $\xi$ is relatively small.
Assertion (1) can also be obtained from other reaction-diffusion models \cite{Is, Ot}.
In contrast, assertion (2) is a characteristic of our results
based on the reaction-diffusion model \eqref{a1} with \eqref{a2}. 
When $|X|$ is constant, we can regard $K$ as a bifurcation parameter.
Let us suppose a situation in which the parameter $K$ increases whereas $|X|$ is constant,
and consider that this situation corresponds to the stage (just) before cell division. 
We can then see that cell polarity oscillations can be observed (just) before cell division,
which is consistent with our naive intuition (cf. \cite{DR}).
Similarly, when $K$ is constant, we can regard $|X|$ as a bifurcation parameter.
We suppose a situation in which the parameter $|X|$ decreases whereas
$K$ is constant. We can then see that cell polarity oscillations occur when
the amount of chemical $X$ decreases. This implies that chemical $X$ is indispensable
for freezing cell polarity.

In simple terms, mathematical models for studying cell polarity oscillations
are classified into two types, i.e., ODE models and reaction-diffusion models.
In general, ODE models
are based on concrete molecular species and networks, and 
explicitly or implicitly assume the existence of the front and back of a cell.  
Such models cannot represent cell polarity oscillations with the non-reversal of polarity,
as observed in cells in a developing epithelial tissue \cite{DR}.
In contrast, the reaction-diffusion model \eqref{a1} with \eqref{a2} can 
represent cell polarity oscillations with both the non-reversal and reversal of polarity. 
Moreover, we note that it does not assume a priori the existence of the front and back of
a cell. 

The Gierer-Meinhardt model \cite{GM2} seems to be the first reaction-diffusion system
in history, which can be used to study spatiotemporal patterns 
consisting of (many) localized unimodal patterns. 
This model has been considered as a reference for studying cell polarity, 
and \cite{GM2} noticed that 
oscillations are necessary for the reorientation of polarity. However, they did not give 
any concrete information concerning the oscillations. More than twenty years later, \cite{Mei}
proposed an ODE system consisting of many components (cells), 
which can represent cell polarity oscillations with the periodic reversal of polarity.
Consequently, as shown in a review paper \cite{JE}, 
reaction-diffusion models have been considered to be undesirable for representing the dynamic maintenance of cell polarity such as cell polarity oscillations.
Our result suggests a counter-example for such a common view, i.e., 
\eqref{a1} with \eqref{a2}
can be a simple reaction-diffusion model for studying cell polarity oscillations.

A reaction-diffusion model of the Rho GTPases (Rac, Cdc42, RhoA),
which are key regulators for cell polarity, was also proposed by \cite{Ot}.  
Moreover, \cite{Ot}
compared this model with reaction-diffusion models proposed by \cite{NSL, SN2},
which are based on a phosphoinositide cycle that recruits and activates the
Rho GTPases (Rac, Cdc42). Consequently, 
the dynamics of these models share common properties, which 
can be interpreted as follows: (1) the spontaneous formation of cell polarity,
(2) the uniqueness of polarity determining the axis of a cell,
(3) the polarized sensitivity wherein a polarized cell responds to a change in
the extracellular signal gradient. 
Although these models based on the concrete 
molecular species and networks are considerably realistic, they consist of many components
and have complicated nonlinear terms, and hence it is difficult to understand
a mechanism for generating the dynamics with three properties. Therefore,
\cite{Ot} provided the concept of reaction-diffusion systems with mass conservation, 
which is given by \eqref{a1}. In fact, the realistic models include
the structure of \eqref{a1}, and it was shown in \cite{Ot}
that \eqref{a1} can exhibit dynamics with 
three properties for appropriate nonlinear terms $f(u, v)$.

The reaction-diffusion system \eqref{a1} with \eqref{a2} is a phenomenological model,
which may not explicitly correspond to concrete molecular species and networks.
Therefore, at present, our results may not be applicable to practical biological problems.
However, as was mentioned above, the reaction-diffusion system \eqref{a1}
with \eqref{a2} can exhibit the dynamics observed in 
the realistic models proposed by \cite{NSL, Ot, SN2}.
Hence, it is important to explore a parameter region where
these realistic models exhibit cell polarity oscillations. Once we find such a region, 
we can propose physical characteristics or parameters related to mechanisms
for switching cell polarity, and can therefore compare our results with those in \cite{TW}. 
Moreover, we might propose experimentally testable hypotheses concerning 
cell polarity oscillations with the aid of biologists. 

Since our results were obtained using 
a specific reaction-diffusion system with bistable nonlinearity and mass conservation,
it is also important 
to investigate conditions in which general reaction-diffusion systems 
with bistable nonlinearity and mass conservation can generate
cell polarity oscillations. This investigation will consist of three steps:
(1) investigating conditions in which \eqref{a1} with bistable nonlinearity has a localized unimodal 
stationary pattern, (2) investigating conditions in which the unimodal pattern 
becomes unstable through a Hopf bifurcation with respect to 
the ratio of the diffusion coefficients by a normal form analysis, as presented in \cite{Ok},
(3) investigating conditions in which oscillatory patterns bifurcating from the 
unimodal pattern undergo a period doubling bifurcation by a numerical bifurcation analysis, 
as presented herein. We expect that the generalization of
nonlinear terms will provide a viewpoint for exploring
biological phenomena, which can be studied by reaction-diffusion systems with bistable nonlinearity and mass conservation.

\vspace{1cm}

{\bf Acknowledgments.} 
The authors would like to express their appreciation to 
the referees for their useful suggestions and comments, which have improved the 
original manuscript.
The first, second, and third authors were supported 
in part by the JSPS KAKENHI 16K05273, 17K14237, 19H01805, respectively.

\vspace{1cm}

\no
{\bf Appendix}  \ The calculations of the right hand sides of \eqref{c5X} and \eqref{c5Y}

\vspace{1cm}

Noting \eqref{b11}, \eqref{c0}, \eqref{c3} and 
$$
\cos ( \dis\frac{2\pi (z + \ell) }{K} ) - \cos ( \dis\frac{2\pi (z - \ell) }{K} )
= -2 \sin  ( \dis\frac{2\pi z }{K} ) \sin  ( \dis\frac{2\pi  \ell }{K} ), 
$$
a direct calculation with the aid of Mathematica shows
$$
\begin{array}{l}
J(\ell) = - \dis\int_0^{K/2} a_1 ( \bar{u} (z) + \bar{v}(z) ) w(z) \{ a_E( z+ \ell) - a_E( z -\ell) \} dz
\\[3ex]
\ \ \ = - \dis\frac{ 4 a_1 b p_0^2 A}{d_2 - d_1} \sin  ( \dis\frac{2\pi  \ell }{K} ) 
\dis\int_0^{\infty} \text{sech}^4(bz) \text{tanh}(bz) \sin  ( \dis\frac{2\pi z }{K} ) dz
+ O( Ke^{-bK})
\\[3ex]
\ \ \ = -
\dis\frac{ 4 a_1 b p_0^2 A}{d_2 - d_1} \cdot \dis\frac{ \pi^3(b^2 K^2 + \pi^2)}{3b^5 K^4} 
\cdot \text{cosech}(\frac{\pi^2}{bK})
\sin  ( \dis\frac{2\pi  \ell }{K} ) + O( Ke^{-bK}).
\end{array}
$$
Similarly, we have
$$
\begin{array}{l}
\dis\int_0^{K/2} w^2(z) dz
=
\frac{4b^2 p_0^2}{(d_2-d_1)^2} \int_0^{\infty} \text{sech}^4(bz)\text{tanh}^2(bz) dz 
+ O( Ke^{-bK})
\\[3ex]
\ \ \
= \dis\frac{ 8b p_0^2}{15(d_2- d_1)^2} + O( Ke^{-bK}).
\end{array}
$$
Since
$(\text{sech}^2(bx))' = -2b \, \text{sech}^2(bx) \text{tanh}(bx)$, 
$\text{tanh}(bx)' = b \, \text{sech}^2(bx)$, 
and $(\varphi_2^*)'' = h_2$, 
using \eqref{c550}, $\varphi_2^*(0) = \varphi_2^*(K/2) = 0$, and the integration by parts twice, we have
$$
\begin{array}{l}
2(d_2 -d_1) \dis\int_0^{K/2} w(z) \varphi_2^*(z) dz
= 2 p_0 \dis\int_0^{K/2} (\text{sech}^2(bz))' \varphi_2^*(z) dz
\\[3ex]
\ \ \ = - 2 p_0 \dis\int_0^{K/2} \text{sech}^2(bz) ({\varphi_2^*}(z))' dz
=  -\dis\frac{2 p_0}{b} \int_0^{K/2} (\text{tanh}(bz))' ({\varphi_2^*}(z))' dz
\\[3ex]
\ \ \ =  - \dis\frac{2 p_0}{b} \left\{ \frac{d\varphi_2^*}{dx}(K/2)  -
\int_0^{K/2} \text{tanh}(bz) ({\varphi_2^*}(z))'' dz \right\}
\\[3ex]
\ \ \ = - \dis\frac{2 p_0}{b} \left\{  \frac{2}{K}\int_0^{K/2} x h_2(x) dx
- \int_0^{K/2} \text{tanh}(bz) h_2(z) dz \right\},
\end{array}
$$
where we abbreviate error terms denoted by $O(Ke^{-bK})$.
Moreover, a direct calculation with the aid of Mathematica shows
$$
\begin{array}{l}
\dis\int_0^{K/2} \text{tanh}(bz) h_2(z) dz 
=
- \dis\frac{2 a_1 b p_0 }{d_2 (d_2 -d_1)} \Big{\{}
3 a_2 \dis\int_0^{\infty}  \text{sech}^4(bz)\text{tanh}^2 (bz) dz 
\\[3ex]
\ \ \  +  \, p_0^2 \dis\int_0^{\infty} \text{sech}^6(bz)\text{tanh}^2(bz) dx 
- a_2 \dis\int_0^{\infty}   \text{sech}^2(bz)\text{tanh}^2(bz) dz 
\Big{\}}  + O( Ke^{-bK})
\\[3ex]
\ \ \ =
- \dis\frac{2 a_1 b p_0 }{d_2 (d_2 -d_1)} \Big{\{}
3 a_2 \cdot \dis\frac{2}{15b} +  \, p_0^2 \cdot \dis\frac{8}{105b}
- a_2 \cdot \dis\frac{1}{3b}
\Big{\}}  + O( Ke^{-bK})
\\[3ex]
\ \ \ =
- \dis\frac{2 a_1 p_0 }{d_2 (d_2 -d_1)} \Big{\{}
\dis\frac{a_2}{15} +  \dis\frac{8 p_0^2}{105}
\Big{\}}  + O( Ke^{-bK}).
\end{array}
$$
Therefore, we have
$$
\begin{array}{l}
\lg S_z, \Phi^* \rg = -2d_2 \dis\int_0^{K/2} w^2(z) dz + 2(d_2 -d_1) \dis\int_0^{K/2} w(z) \varphi_2^*(z) dz 
\\[3ex]
= 
- \dis\frac{ 4p_0^2}{15(d_2 -d_1)} 
\left\{
\dis\frac{4b d_2}{d_2-d_1} + \dis\frac{a_1 a_2}{b d_2} 
+ \dis\frac{8 a_1 p_0^2}{b d_2}\Big{(} \dis\frac{1}{7} - \dis\frac{1}{3Kb} \Big{)}
\right\}  + O(Ke^{-bK}) < 0.
\end{array}
$$

\end{document}